\documentclass[useAMS]{mn2e}
\usepackage{graphicx}

\begin{document}

\title[Simulations of central AGN evolution in dynamic galaxy clusters]{Swimming against the current: Simulations of central AGN evolution in dynamic galaxy clusters}

\author[B. J. Morsony et al.]{Brian J. Morsony,$^{1}$\thanks{Email:morsony@astro.wisc.edu} Sebastian Heinz,$^1$ 
Marcus Br\"uggen,$^2$ and Mateusz Ruszkowski$^{3,4}$\\
$^1${Department of Astronomy, University of
Wisconsin-Madison, 3321 Sterling Hall, 475 N. Charter Street, Madison WI
53706-1582, USA}\\
$^2${Jacobs University Bremen, Campus Ring 1, 28759 Bremen, Germany}\\
$^3${Department of Astronomy, The University of Michigan, 500 Church Street, Ann Arbor, MI 48109, USA}\\
$^4${The Michigan Center for Theoretical Physics, 3444 Randall Lab, 450 Church St, Ann Arbor, MI 48109, USA}}


\maketitle

\label{firstpage}

%
%
%
%
%
%

\begin{abstract}

We present a series of three-dimensional hydrodynamical simulations of
central AGN driven jets in a dynamic, cosmologically evolved galaxy
cluster.  Extending previous work, we study jet powers ranging from $L_{\rm jet}=10^{44}$~erg/s
to $L_{\rm jet}=10^{46}$~erg/s and in duration from $30$~Myr to
$200$~Myr.  We find that large-scale motions of cluster gas disrupt
the AGN jets, causing energy to be distributed throughout the centre
of the cluster, rather than confined to a narrow angle around the jet
axis.  Disruption of the jet also leads to the appearance of multiple
disconnected X-ray bubbles from a long-duration AGN with a constant
luminosity.  This implies that observations of multiple bubbles in a
cluster are not necessarily an expression of the AGN duty cycle.  We
find that the ``sphere of influence'' of the AGN, the radial scale within which the cluster is strongly
affected by the jet, scales as $R \propto L_{jet}^{1/3}$.
Increasing the duration of AGN activity does not increase the radius
affected by the AGN significantly, but does change the magnitude of
the AGN's effects.  How an AGN delivers energy to a cluster will
determine where that energy is deposited: a high luminosity is needed
to heat material outside the core of the cluster, while a
low-luminosity, long-duration AGN is more efficient at heating the
inner few tens of kpc.

\end{abstract}


\section{Introduction}

The X-ray emitting gas in the centre of many galaxy clusters has a
cooling time less than the Hubble time.  Supported by large central
concentrations in the surface brightness profile, early models of
cluster evolution posited that cooling gas is funneled onto the central
galaxy in the form of a cooling flow, ultimately resulting in a
cooling catastrophe as more and more gas condenses out of the hot
phase (Fabian 1994).  
However, high-resolution spectra of clusters show a
lack of cool gas below a temperature floor of about one third of the
virial temperature ($\sim2$~keV) and a lack of star formation at the
levels expected for the early estimates of cooling rates in excess of
100 solar masses per year in powerful cooling flows (e.g., Peterson et
al. 2001; Fabian et al. 2001; Kaastra et al. 2001; Tamura et
al. 2001).  The most natural interpretation is that some source of
heat is preventing the gas from cooling.  The heat source must respond
to the cooling rate of the cluster gas in such a way that heating and
cooling are balanced, on average, in a self-regulating feedback cycle.

Many different mechanisms for heating in clusters have been proposed,
including AGN outflows, thermal conduction, dynamical friction,
gravitational heating, cosmic rays and preheating, each either alone
or in combination (see Conroy \& Ostriker 2008; Brighenti \& Mathews
2002 for reviews).  However, recent studies have found evidence for
AGN activity in the central galaxy of nearly all cooling-core clusters
(Burns 1990; Mittal et al. 2009).  

The now widely accepted picture is that radio-loud AGN inflate bubbles
of under-dense, relativistic plasma which displaces the local ICM.
The plasma bubbles appear as dark cavities in X-ray observations (see,
for example, B\^irzan et al. 2004; Dunn \& Fabian 2004; Rafferty et
al. 2006).  While the presence of cavities is ubiquitous, the longer
term evolution of the plasma, and the direct effects on the gas are
still uncertain and a topic of ongoing work (e.g. Reynolds et al. 2002; 
Ruszkowski et al. 2004b; Zanni et al. 2005; Gaibler et al. 2009).

Although it is possible that AGN outflows are not solely or primarily
responsible for cluster feedback, understanding how AGN interact with
the cluster environment is still necessary to understand how clusters
evolve.


Interactions with the cluster environment can have an effect on other 
types of cluster heating.
Recent work by Parrish et al. (2010) and Ruszkowski \& Oh (2010) has found that turbulence in the 
cluster above a threshold value can 
suppress the heat-flux-driven buoyancy instability which would other 
wise inhibit thermal conduction.
Turbulence can, therefore, allow efficient thermal conduction to 
occur, so it is important to understand the velocity structure of the 
cluster and how AGN alter that structure.


\subsection{AGN simulations}

The details of how an AGN delivers energy to the cluster are still
largely unknown.  Simulations of AGN jets or hot, under-dense bubble
in idealized, spherically symmetric clusters have found that energy is
confined to a narrow angle around the jet axis, rather than being
spread throughout the cluster (e.g. Churazov et al. 2001; Reynolds et al. 2001; 
Saxton et al. 2001; Br\"uggen et al. 2002; Br\"uggen \& Kaiser 2002; 
Reynolds et al. 2002; Ruszkowski et al. 2004a; Dalla Vecchia et al. 2004; 
Omma et al. 2004; Omma \& Binney 2004; O'Neil et al. 2009; O'Neil \& Jones 2010).
Turbulence produced by RT instabilities has been shown to help in
distributing energy and in producing a self-regulated AGN jet 
(Scannapieco \& Br\"uggen 2008, Br\"uggen \& Scannapieco 2009).

However, a simulation carried out by Heinz et al. (2006) found that in
a realistic, cosmologically evolved cluster the motion of cluster gas
effectively distributed the effects of the AGN over a wide angle.  
There is also observational evidence for the misalignment of radio jets 
and X-ray cavities in Abell 4059, presumably due to bulk motions in the 
ICM (Heinz et al. 2002).
MHD simulations of clusters with AGN and turbulence from star 
formation (Falceta-Gon\c{c}laves et al. 2010b) were able to 
distribute AGN energy isotropically and produce a filamentary 
structure similar to the observed in Perseus.


An alternate means of distributing energy over a wide angle is to 
change the properties of the AGN rather than the cluster.
Two-dimensional simulations in spherically symmetric clusters of slow, 
wide opening angle jets (Sternberg et al. 2007) or narrow, rapidly precessing 
jets with a wide precession angle (Sternberg \& Soker, 2008) have been 
able to effectively distribute energy in the cluster core.  
Three-dimensional simulations by Falceta-Gon\c{c}laves et al. (2010a) 
of slowly precessing jets were also able to produce an approximately 
isotropic energy distribution.  
These simulations were also able to create multiple bubbles if a large 
precession angle ($\approx 60\degr$) was used.


In this paper, we directly investigate the role of cluster weather and
jet power, as well as the injection history, on the evolution of radio
lobes and X-ray cavities in the ICM.

\subsection{Jet intermittency}
One critical question concerning the effect of jet activity is what
the jet duty cycle is, i.e., what fraction of the time a black hole at
the centre of a cool core cluster is actively driving a jet.  
The average jet power is often used to estimate the amount of AGN heating.
However, the heating from a short, powerful period of AGN activity could 
be different from a long-lived, low-luminosity AGN that injects the same 
amount of total power.

Observations of multiple radially segregated sets of cavities in
several clusters have been used to argue that jet activity is
intermittent.  Examples include Perseus (B\"{o}hringer et al. 1993; Fabian et
al. 2000, 2003, 2006), Hydra A (Wise et al. 2007), Virgo (Forman et
al. 2007), and Abell 262 (Clarke et al. 2009).  The duty-cycle
inferred from these observations has been taken as evidence for
self-regulation between cooling of the central cluster gas and AGN
activity on timescales of tens of millions of years.  Similar
arguments have been made on the basis of sound and shock waves
observed in deep Chandra observations of nearby clusters (Virgo,
Perseus), which associate an AGN outburst with every ripple.

As we will argue, the underlying assumption that surface brightness
features can be associated one-to-one with activity in the central
engine is likely overly optimistic.  Sternberg \& Soker (2009) have 
found that multiple sound waves can be excited by a single episode 
of bubble formation.  
Falceta-Gon\c{c}laves et al. (2010a) were able to create multiple 
bubbles in simulations of a precessing AGN jet with a large 
precession angle.
Cluster weather and the dynamic
nature of the evolution of radio lobes can introduce features very
similar to those observed in nearby clusters even in the case of
continuously powered jets with no actual modulation of the jet power.

In this paper, we present the results of hydrodynamical simulations of
AGN jets in a realistic galaxy cluster with a variety of AGN
properties, focusing on how the interaction of the AGN with a dynamic
ICM affects the evolution of the morphology and energy distribution of
the AGN.  The paper is arranged as follows: Section 2 describes the
setup and technical details of our simulations, Section 3 presents the
results and discusses the implications for cluster heating
and observations, Section 4 presents a analytic toy model to estimate 
the characteristic timescale for an individual bubble form and break 
off from the jet, and Section 5 presents a summary of our results and
conclusions.

\section{Technical Description}

\subsection{Code description}

Simulations are carried out using the FLASH 2.4 hydrodynamics code
(Fryxell et al. 2000), which is a modular, block-structure adaptive
mesh code.  It solves the Riemann problem on a three-dimensional
Cartesian grid using the piecewise-parabolic method.  Our simulations
include gravity from $7\times10^{5}$ dark matter particles, advanced
using a cosmological variable-time-step leapfrog method.  Gravity is
computed by solving Poisson's equation with a multigrid method using
isolated boundary conditions.  Radiative cooling and star formation
are not included in our simulations, but for the relatively short time
simulated (200 Myr) these 
should not have a significant effect on the large scale evolution of the cluster and
can be neglected.  
Gas is modeled as having a uniform adiabatic index of $\gamma = 5/3$.

\subsection{Initial Conditions}

Our simulations are carried out in a dynamic 
cluster that has been extracted from a cosmological SPH simulation.  
The cluster is based on a rerun of the S2 cluster in
Springel et al. 2001 and is identical to the cluster used in Heinz et
al. (2006), Br\"uggen et al. (2007) and Heinz et al. (2010).  
Radiative cooling and star formation were included in the SPH simulation 
used to construct the cluster we use for our initial setup.
The cluster is a massive, X-ray bright cluster with a mass
of $M \approx 7\times10^{14}$~M$_{\sun}$ and a central temperature of
$6$~keV.  There is a net circular motion in the cluster centre as well
as large directional motions spanning 100's of kpc farther out, and a
dynamically-induced cold front.  This cluster is chosen specifically
because it is not relaxed, allowing us to determine how the motion of
cluster gas affects the evolution of the AGN jet.

Disturbed clusters are common in cosmological simulations (e.g. Burns et al. 2008) 
and in real clusters based on X-ray surface brightness analysis (e.g. Schuecker et al. 2001).
Little is currently known about the detailed velocity structure of gas in galaxy 
clusters, but future high spectral resolution observations with the 
International X-ray Observatory ({\em IXO}) may be able to provide this information.

For comparison purposes, we have also carried out a control simulation 
in a hydrostatic cluster.  The hydrostatic cluster is set up
such that the radial gas density and gravitational profile are similar
to the average profiles of our realistic cluster.  The pressure is
then set such that the cluster is in hydrostatic equilibrium
everywhere, and the gas is stationary everywhere.  The gas density
profiles for this cluster is fit by the sum of two exponential
functions with the form $\rho(r) = a_1e^{a_2r^{a_3}} +
a_4e^{a_5r^{a_6}}$ where $r$ is the radius in cm from
the cluster centre and $\rho$ is the mass density in g/cm$^3$.  
The values of $a_i$ can be found in Table~\ref{table_hydro}.
The gravitational potential profile is fit by a logarithm of the form
$g_{pot}(r) = b_1 log(r/b_2) - b_3$ where $r$ is the radius in cm from the cluster
centre and $g_{pot}$ is the gravitational potential.  The values of $b_i$ can be found in Table~\ref{table_hydro}.

\begin{table}
\caption{Hydrostatic Cluster Parameters \label{table_hydro}}
\begin{tabular}{@{}ll}
\hline
{Parameter} & {Value}\\
\hline

$a_1$ & $9.32\times10^{-25}$ \\
$a_2$ & $-1.86\times10^{-12}$ \\
$a_3$ & $5.30\times10^{-1}$ \\
$a_4$ & $2.48\times10^{-27}$ \\
$a_5$ & $-3.98\times10^{-32}$ \\
$a_6$ & $1.29$ \\
$b_1$ & $3.56\times10^{16}$ \\
$b_1$ & $8.39\times10^{21}$ \\
$b_1$ & $1.32\times10^{17}$ \\

\hline
\end{tabular}
\end{table}

\subsection{Jet injection}

The computational domain consists of a 2.8 Mpc$^{3}$ box centred on the cluster.  
The maximum resolution is $\sim22$ pc near the
jet nozzle.  Simulations are carried out by placing two oppositely
directed jets at the centre of the gravitational potential of the
cluster.  The jet nozzle is modeled as two circular disks back-to-back
with inflow boundaries, resolved by 12 grid elements in diameter.  The
nozzle faces obey inflow boundary conditions fixed by the mass-,
momentum- and energy fluxes of the jet.  Unresolved dynamical
instabilities near the base of the jet are modeled by imposing a
random-walk jitter on jet axis confined to a $20{\degr}$ half-opening
angle.  
This is necessary to model the `dentist's drill' effect of Scheuer (1982). 
Jet material is injected with a velocity of $v_{\rm jet} =
3\times10^{9}$~cm/s and an internal Mach number of 10 in all
simulations.

The simulations we present vary in AGN luminosity and duration.
Simulations 44, 45 and 46 have AGN luminosities of $10^{44}$~erg/s,
$10^{45}$~erg/s and $10^{46}$~erg/s, respectively, and all 3
simulations have an AGN active for the first 30 Myr
of the simulation.  Simulations 45L and 45C have a luminosity of
$10^{45}$~erg/s, the same as simulation 45, but the duration of AGN
activity is increased to 90 Myr for
simulation 45L and is continuous for the entire 200 Myr run time for
simulation 45C.  We also carry out a control simulation with no AGN
present.  Comparing cluster properties with and without an AGN at the
same time allows us to separate the effects of the AGN from the
dynamical evolution of the cluster gas.  See table~\ref{table_setup}
for details of the setups for each simulation.  All simulations are run
for a total of 200 Myr.

The simulation carried out in the hydrostatic cluster (45S in
table~\ref{table_setup}) is identical to simulation 45C, with an AGN
luminosity of $10^{45}$~erg/s that is on continuously for 200 Myr.

\begin{table}
\caption{Simulation Parameters \label{table_setup}}
\begin{tabular}{@{}llll}
\hline
{Name} & {Luminosity} & {Duration} & {Cluster} \\
\hline

44 & $10^{44}$ erg/s & 30 Myr & Realistic \\
45 & $10^{45}$ erg/s & 30 Myr & Realistic \\
46 & $10^{46}$ erg/s & 30 Myr & Realistic \\
45L & $10^{45}$ erg/s & 90 Myr & Realistic \\
45C & $10^{45}$ erg/s & Continuous & Realistic \\
45S & $10^{45}$ erg/s & Continuous & Hydrostatic \\
Control & $0$ & $0$ & Realistic \\
Hydrostatic Control & $0 $ & $0$ & Hydrostatic \\

\hline
\end{tabular}
\end{table}

\subsection{Visualization}

The simulation output was virtually observed using the publicly
available in-house tool XIM (see Heinz \& Br\"uggen, 2009).  Taking input
grids from numerical hydrodynamic simulations, XIM performs spectral
modeling of thermal emission, including Doppler shifts and ionization
balance, using the APEC database to model the line emission.  It then
performs spectral projection along an arbitrary line--of--sight, PSF
convolution, telescope and detector efficiency, and spectral
convolution with the detector response (using the proper response
files for current and future telescopes).  Finally, it adds sky- and
instrument backgrounds and Poisson counting error.

The code currently does not account for vignetting and uses a
simplified mono-energetic PSF in the case of {\em IXO} simulations.
However, the impact of these limitations on the predictions presented
below should be small.

Our virtual observations use the Chandra ACIS instrument response.  The broad-band
images shown span an energy range from 0.3 keV to 7 keV.  For direct
comparison, the images were placed at the redshift of the Perseus
cluster of z=0.01756.  The exposure time was taken to be 250 ksec.

For the pseudo-synchrotron radio images shown below, we assume
equipartition and a completely tangled magnetic field.  The electrons
are assumed to obey an $E^{-2}$ powerlaw spectrum.
Radio intensity is calculated as $L_{radio} \propto B \times P^{1.75}$ integrated 
along the line of sight, where $B$ is the local fraction of jet material 
and $P$ is local pressure.
All radio images are plotted in arbitrary units, but on the same 
logarithmic scale, covering 4 orders of magnitude.

\section{Results}

\subsection{Morphological description}

\begin{figure*}
\includegraphics[scale=.6]{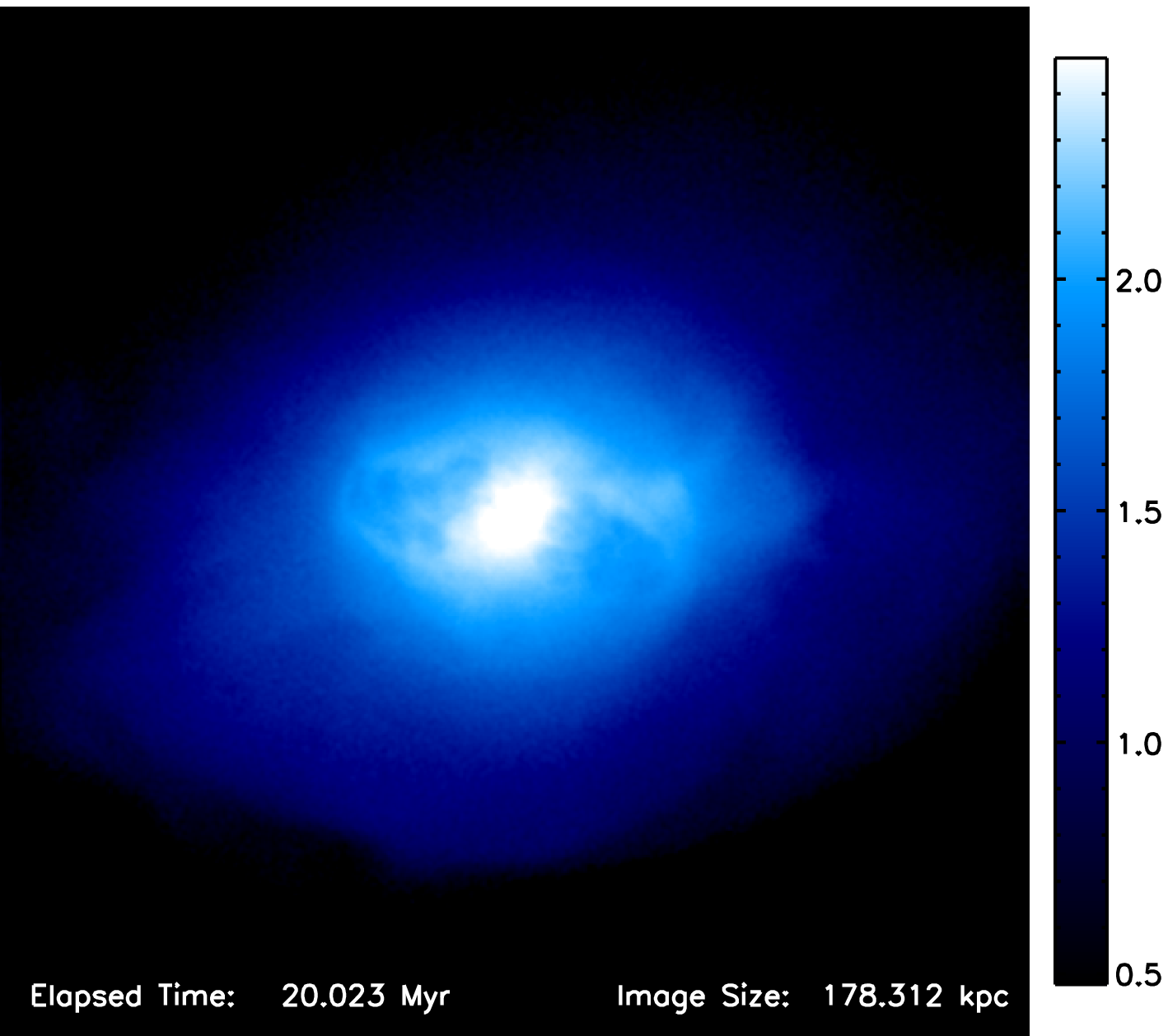}
\includegraphics[scale=.6]{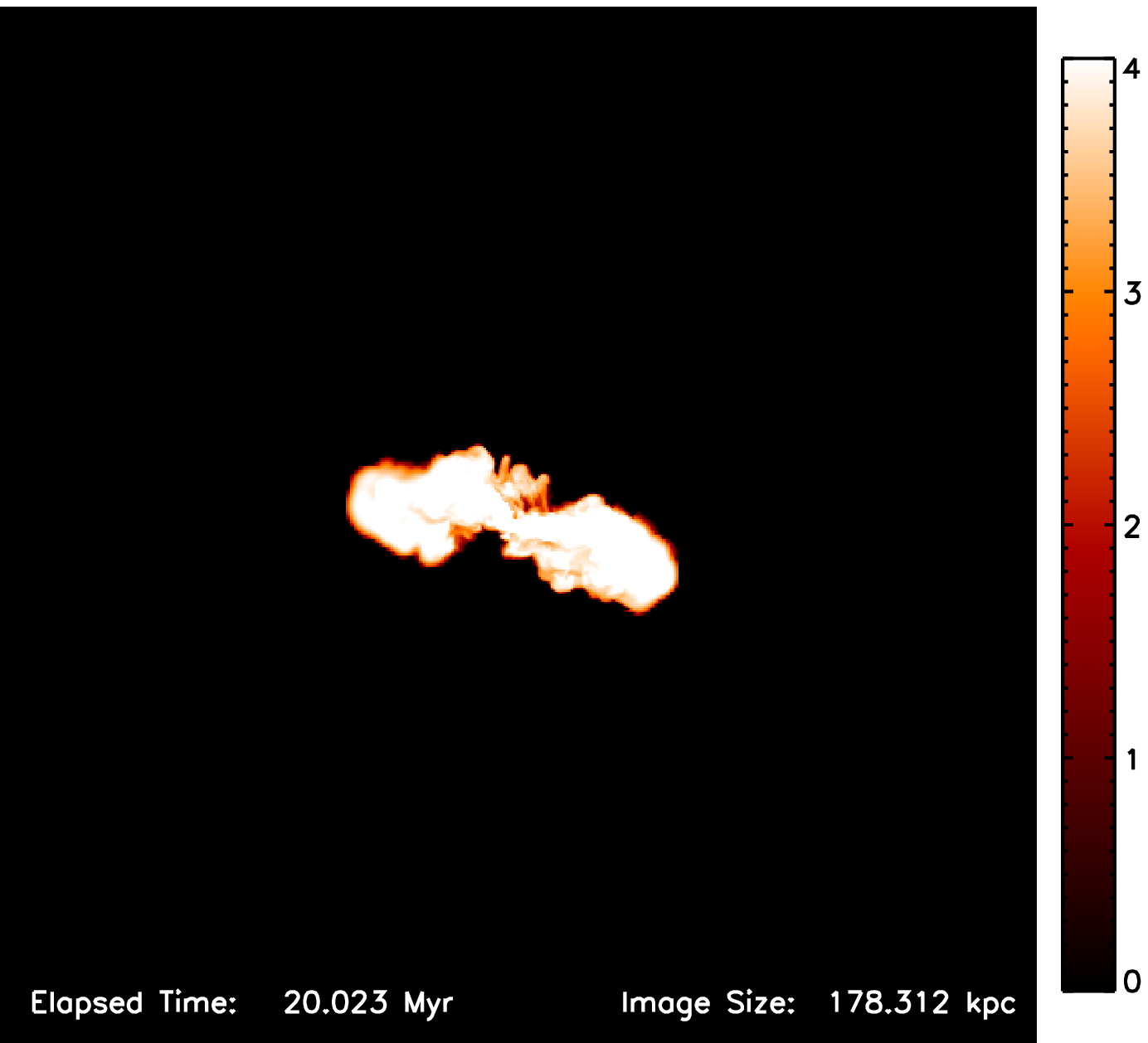}
\caption{
    Synthetic Chandra X-ray image (left, log scale, counts/pixel) and synthetic radio image (right, log scale, arbitrary units) of simulation 45 at 20 Myr.
    The jet has inflated two large bubbles that appear as dark areas in the X-ray image to the left and right of the cluster centre, and as bright radio lobes in the radio image.
}
\label{fig_image_45}
\end{figure*}

The generic features of all three base simulations (44, 45, 46) are
similar.  Figure~\ref{fig_image_45} shows synthetic X-ray and radio
images for simulation 45 after 20 Myr. 

\begin{figure*}
\includegraphics[scale=1.]{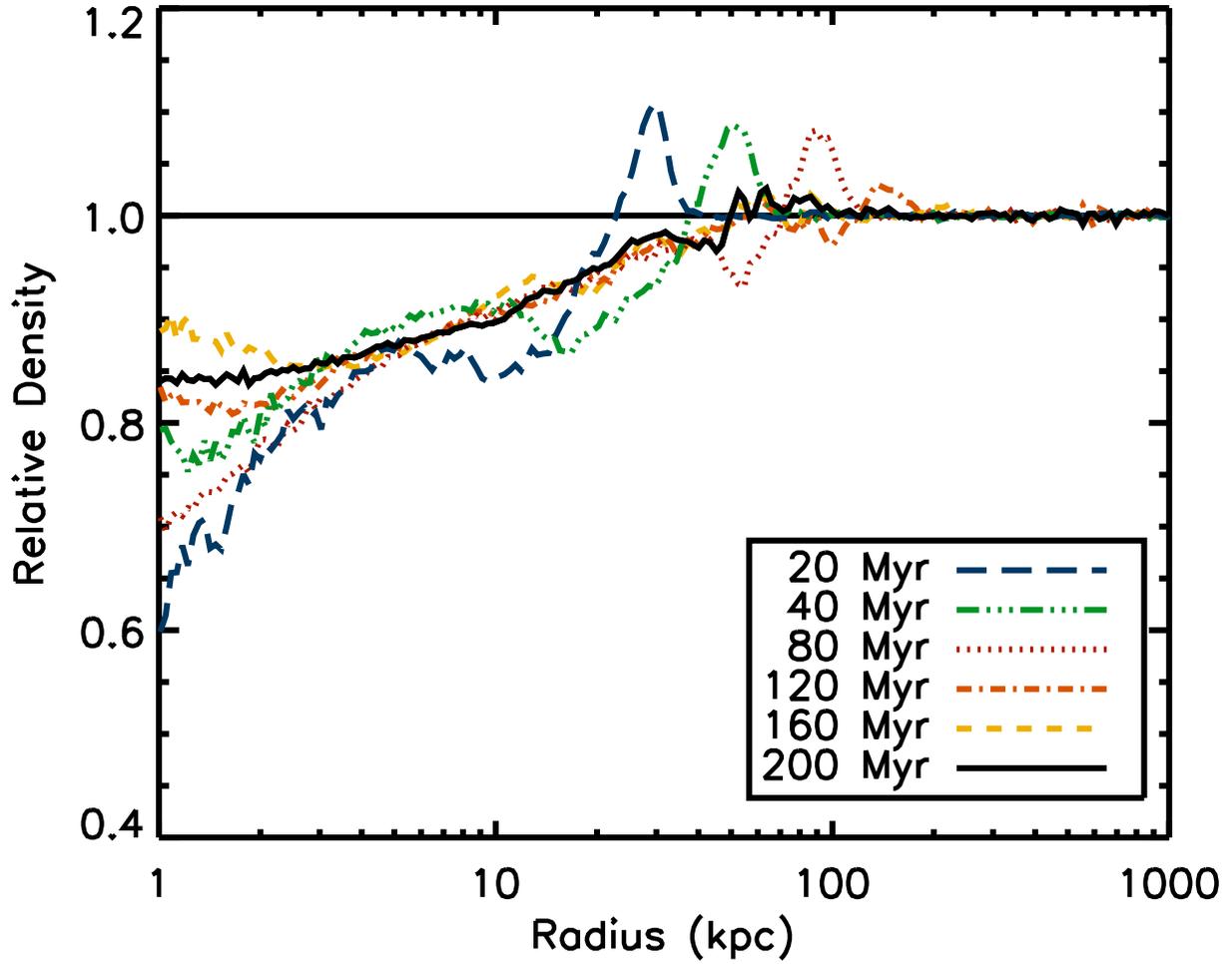}
\caption{
    Density relative to simulation with no AGN vs. radius for simulation 45 from 20 to 200 Myr.
    Although there is some variation, by 120 Myr (90 Myr after AGN turn off) the radius of density decrease remains constant while the over dense region slowly expands.
}
\label{fig_dens_time_45}
\end{figure*}

The AGN jet initially inflates a pair of low-density bubbles which
appear morphologically similar to observations of X-ray bubbles and
radio lobes.  After the AGN turns off, these bubbles quickly
decelerate and reach a fixed radius by about 100 Myr.
Figure~\ref{fig_dens_time_45} shows the average density relative to a
control simulation with no AGN vs. radius for simulation 45 at
different times.  The jet inflates a region of low density in the
centre of the cluster and drives a high density wave outward.  
After the jet shuts off, the under-dense region expands for a while 
but after 120 Myr has a fixed extent of 50~kpc. 
The high-density region decelerates but continues to expand slowly, 
increasing in size from 160~kpc at 120 Myr to 250~kpc at 200 Myr.

\begin{figure*}
\includegraphics[scale=1.]{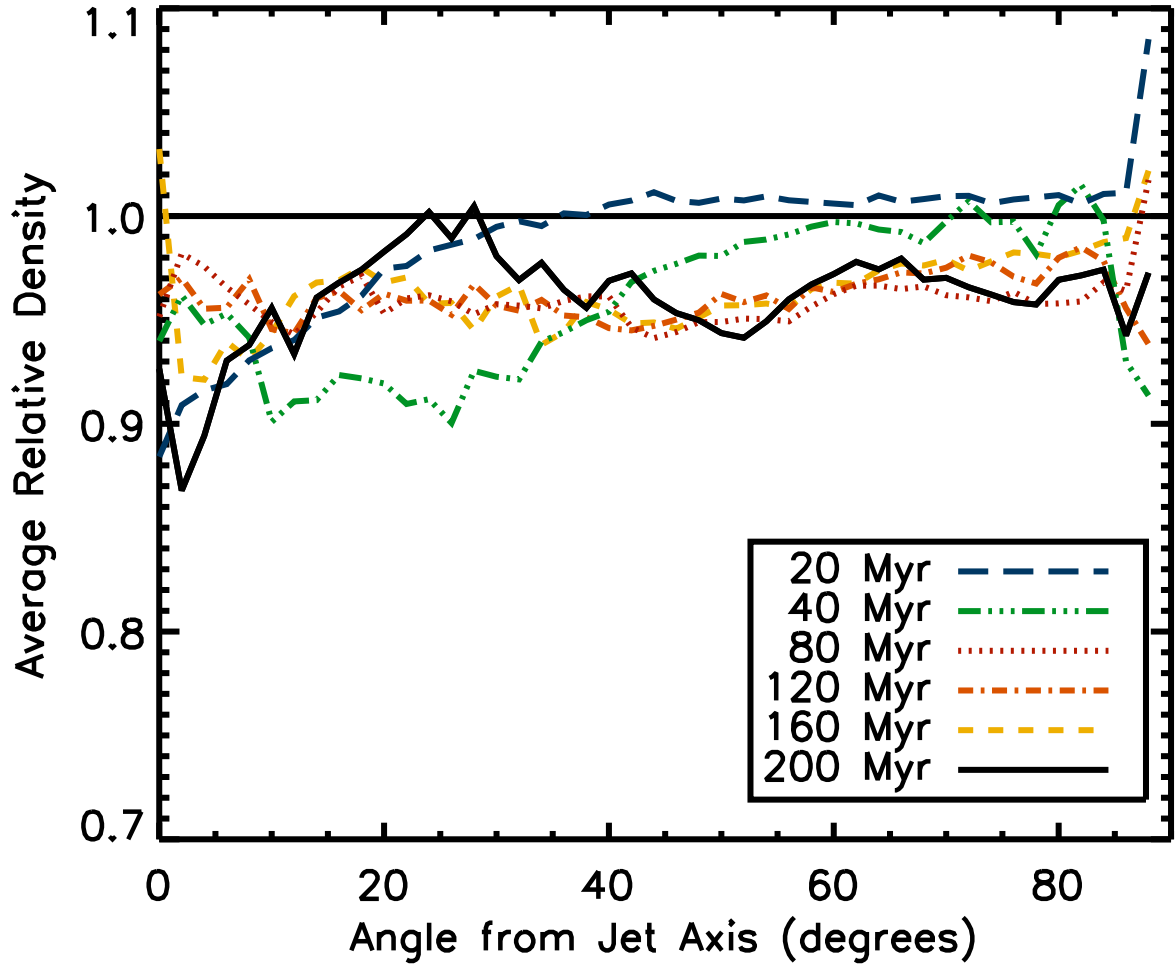}
\caption{
    Average density within 50 kpc relative to simulation with no AGN vs. angle for simulation 45 from 20 to 200 Myr.
    After the AGN turns off at 30 Myr, the effects of the AGN quickly spread over all angles and, by 80 Myr, the density distribution is uniform.
}
\label{fig_dens_time_angle_45}
\end{figure*}

\begin{figure*}
\includegraphics[scale=.4]{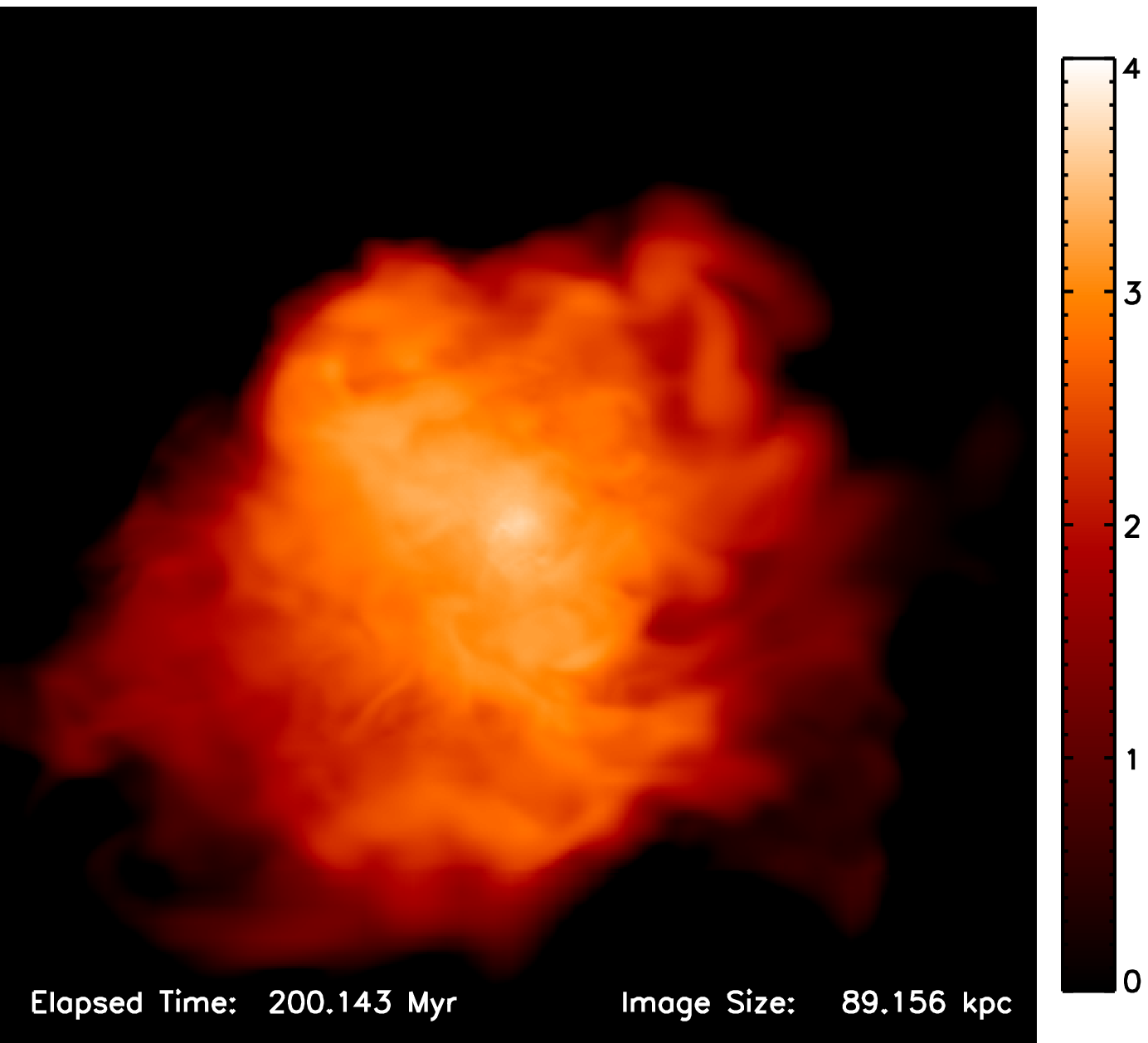}
\includegraphics[scale=.4]{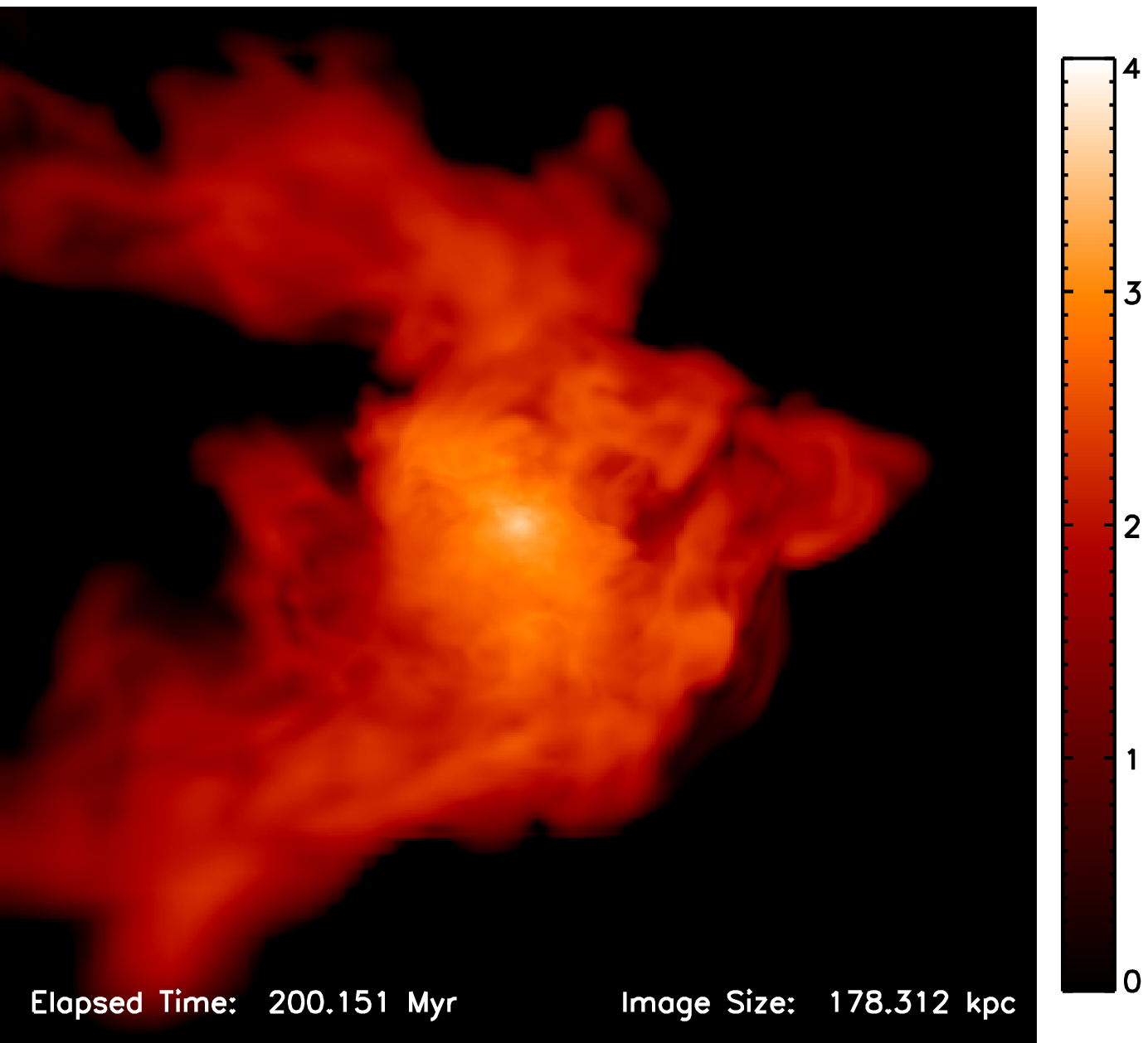}
\includegraphics[scale=.4]{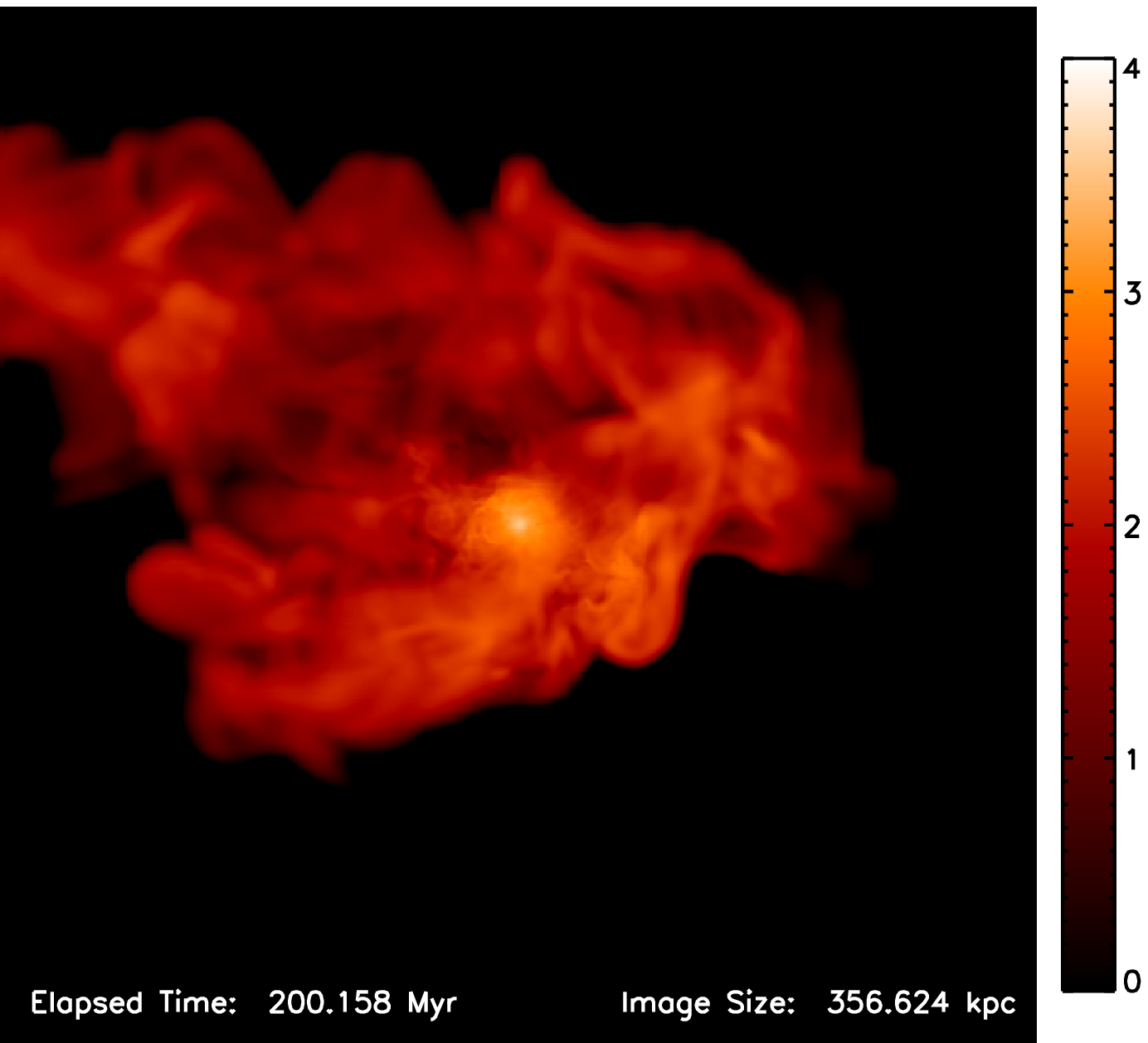}
\caption{
    Comparison of synthetic radio image (log scale, arbitrary units) at 200 Myr for simulations 44, 45 and 46 from left to right.  
    In all cases, jet material is circularized by the motion of gas in the cluster.  
    Note that the scale of the images increases by a factor of two for each image from left to right.
}
\label{fig_image_44_45_46}
\end{figure*}

The evolution at later times is determined mainly by the motion of gas
within the cluster.  Looking at a plot of the angular dependence of
average relative density within 50 kpc of the cluster centre
(Fig.~\ref{fig_dens_time_angle_45}), we see the density decrease is
concentrated within 30 degrees of the jet axis at 20 Myr, but by 80
Myr the decrease is nearly uniform across all angles.  Large-scale
flows within the cluster, in the form of turbulence, rotational
motions and large-scale direction flows, effectively disrupt the
bubbles and distribute them throughout the centre of the cluster.
Figure~\ref{fig_image_44_45_46} shows synthetic radio emission for
simulations 44, 45 and 46 after 200 Myr.  In all cases the jet
material is well spread out and most of the information about the
initial direction of the jet (left to right in the images shown) has
been lost.  However, the radius reached by the jet material increases
with increasing energy.  In Fig.~\ref{fig_image_44_45_46} the size
scale of the radio emission increases by about a factor of two for
each factor of 10 increase in power.

\begin{figure*}
\includegraphics[scale=1.]{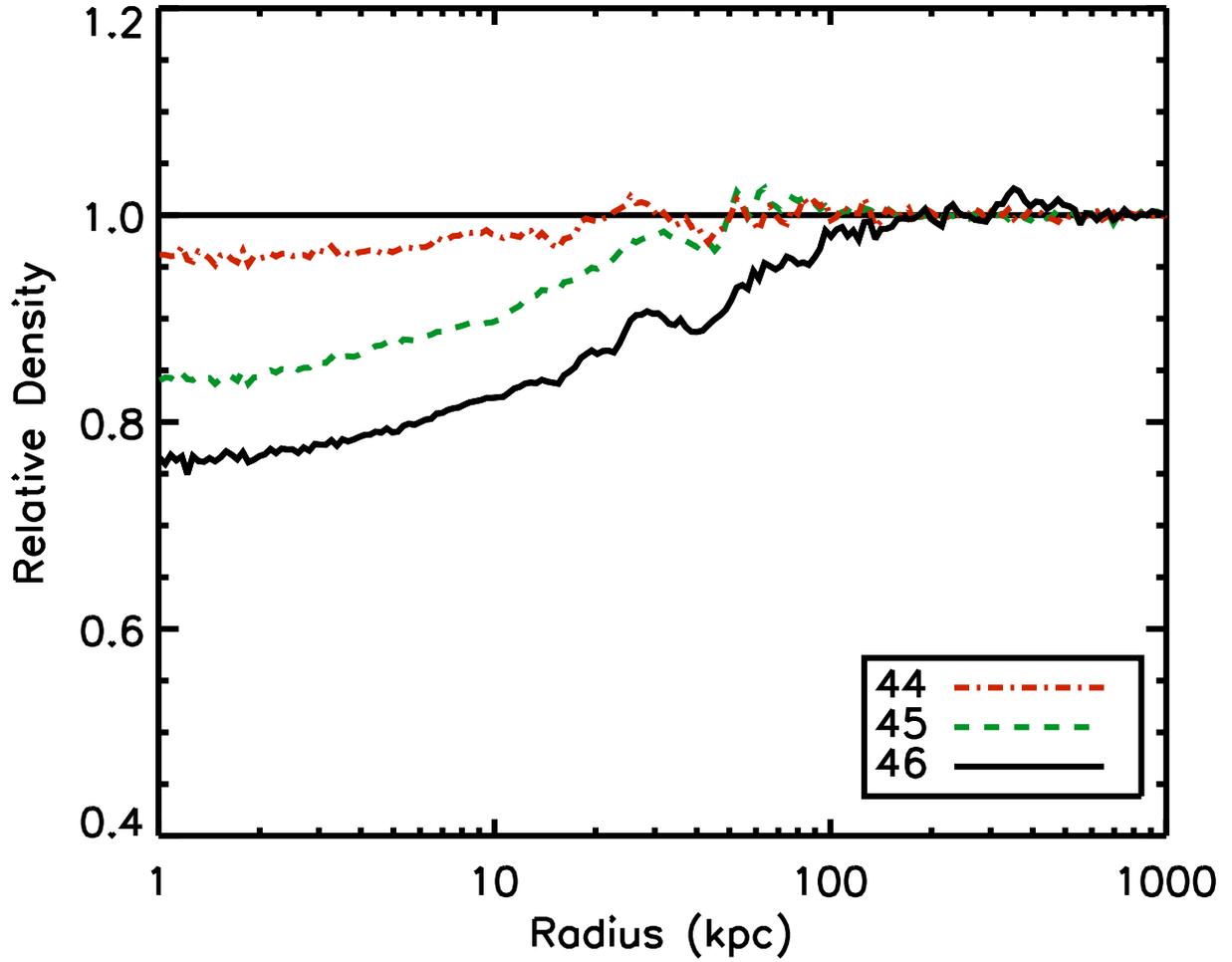}
\caption{
    Density relative to simulation with no AGN vs. radius after 200 Myr for simulations 44, 45 and 46.  
    The radius within which mass had been removed is about 20 kpc, 50 kpc and 100 kpc, respectively.
    The magnitude of density decrease is also larger with increasing AGN luminosity, with 5\%, 15\% and 25\% decreases in the inner 10 kpc.
}
\label{fig_dens_44_45_46}
\end{figure*}

Plots of average density relative to a control simulation with no AGN
vs. radius (Fig.~\ref{fig_dens_44_45_46}) show a decrease in density
in the centre of the cluster out to about 20 kpc, 50 kpc and 100 kpc
for simulations 44, 45 and 46, respectively.  This is consistent with
the volume of density decrease scaling linearly with luminosity (or
radius of density decrease $R \propto L^{1/3}$).

\begin{figure*}
\includegraphics[scale=1.]{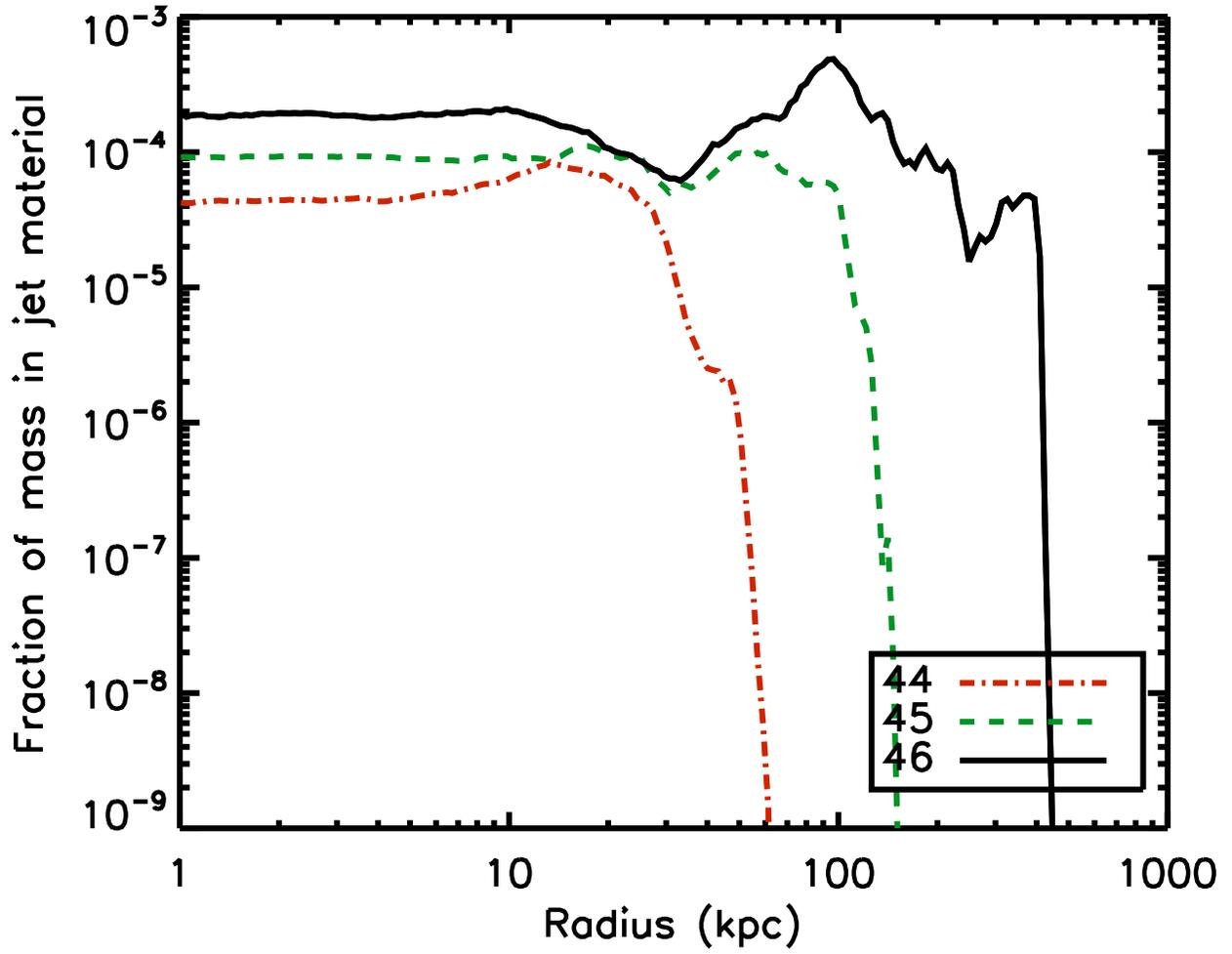}
\caption{
    Fraction of mass comprised of jet material vs. radius after 200 Myr for simulations 44, 45 and 46.  
    The maximum radius that any jet material reaches is about 60 kpc, 150 kpc and 450 kpc, respectively.
}
\label{fig_fraction_44_45_46}
\end{figure*}

There are several ways that the ``radius of AGN influence'' can be defined.
In addition to the radius of density decrease, the maximum radius that 
any jet material reaches and the maximum radius of the high-density wave 
surrounding the AGN can also be used.
The maximum radius that any jet material reaches, corresponding to the 
maximum radius of any synthetic radio emission in Fig.~\ref{fig_image_44_45_46}, 
is about 60~kpc, 150~kpc and 450~kpc after 200 Myr for the 44, 45, and 46 
run, respectively.  
The maximum radius reached by any jet material can also be seen as a sharp 
cutoff in Fig.~\ref{fig_fraction_44_45_46}, which plots the fraction of 
jet material vs. radius.

The maximum radius of the high-density wave corresponds to the maximum radius 
at which there is a density discrepancy in Fig.~\ref{fig_dens_44_45_46}.
The wave extends to about 100 kpc, 250 kpc and 500 kpc at 200 Myr in the 
three simulations.
For any of these three definitions of ``radius of AGN influence'' the 
radius scales with luminosity roughly as $R \propto L^{1/3}$ at any time 
during the simulation.

The magnitude of the density decrease in the cluster also depends on AGN luminosity.
In Fig.~\ref{fig_dens_44_45_46} the density decrease in the inner few
kpc is about 5\%, 15\% and 25\% for simulations 44, 45 and 46.

\begin{figure*}
\includegraphics[scale=1.]{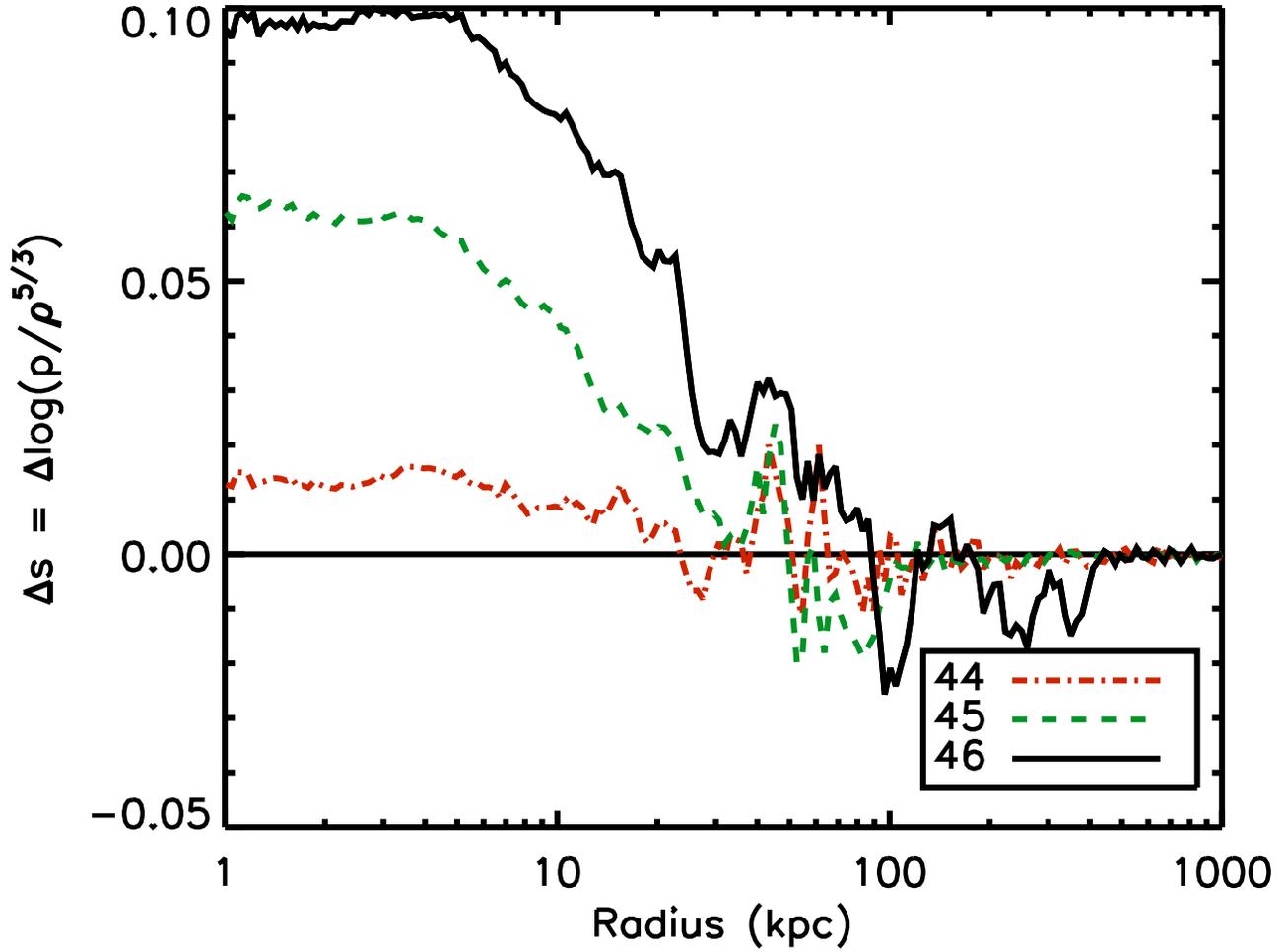}
\caption{
    Average local change in thermal cluster entropy compared to simulation with no AGN vs. radius after 200 Myr for simulations 44, 45 and 46.
    There is an increase in entropy out to about 20 kpc, 50 kpc and 100 kpc, respectively, corresponding to the decrease in density seen in Fig.~\ref{fig_dens_44_45_46}.
}
\label{fig_entropy_44_45_46}
\end{figure*}

Other variables show a similar radial profile to the density plots.
For example, Fig.~\ref{fig_entropy_44_45_46} plots the average change
in entropy for simulations 44, 45 and 46 at 200 Myr.  There is an
increase of entropy in the cluster centre corresponding to the decrease 
in average density within 20 kpc, 50 kpc and 100 kpc, respectively.  
The relative change in entropy also increases with luminosity, with values 
of $\Delta s \approx .01$, $.06$ and $.1$ in the inner few kpc.  
This indicates that most of the heating of the cluster gas is taking place near the
centre of the cluster and in the region where the density has
decreased most strongly.
Outside the radius of density decrease the change in entropy is quite complex. 
Low entropy gas has been lifted out of the cluster centre and mixed with higher 
entropy gas while the AGN has prevented some gas from falling in as far as it 
would have otherwise.
There are changes in entropy out to the shock radius, but they do not follow a 
consistent pattern.

\subsection{Dependence on duty cycle}

\begin{figure*}
\includegraphics[scale=.6]{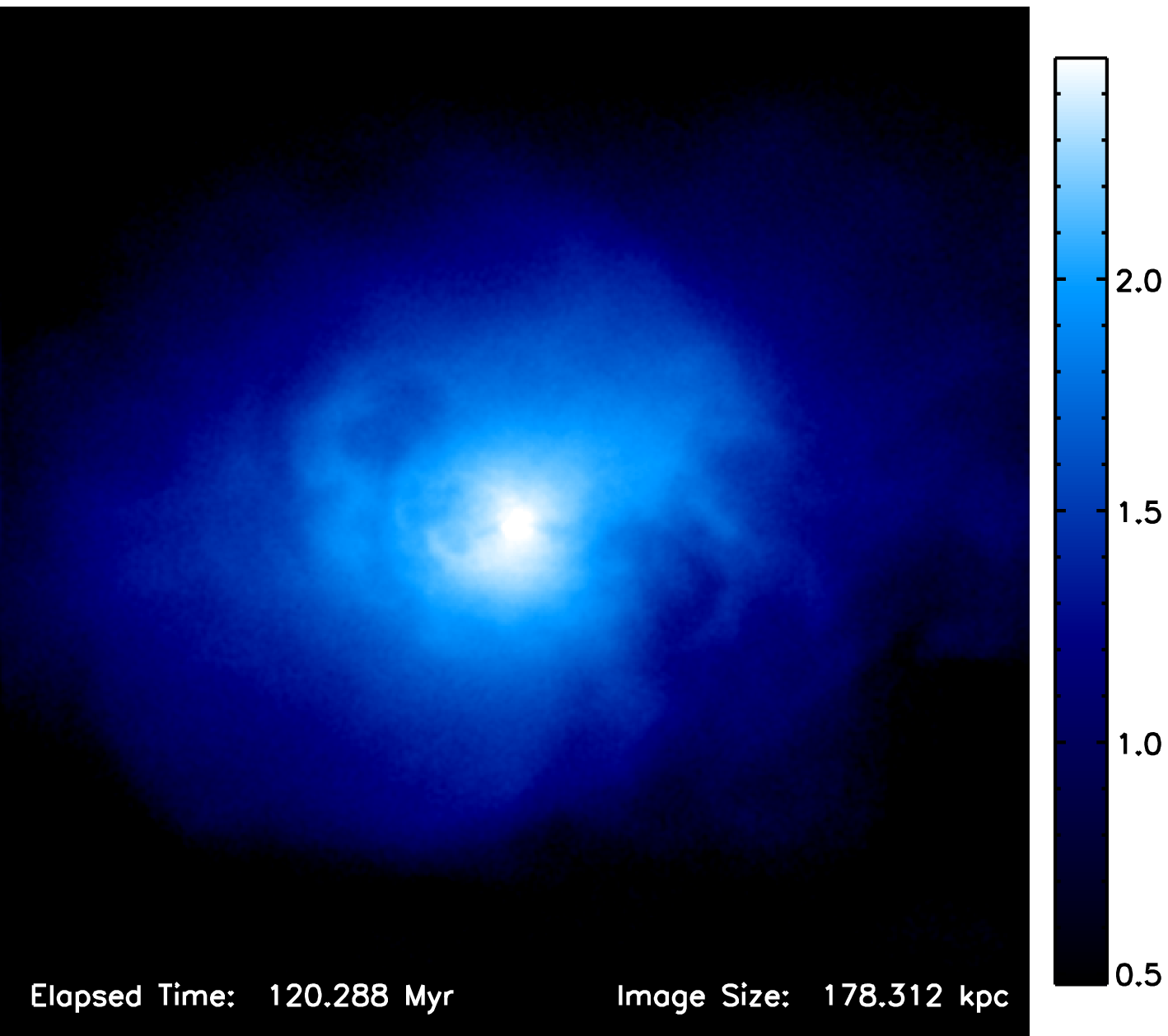}
\includegraphics[scale=.6]{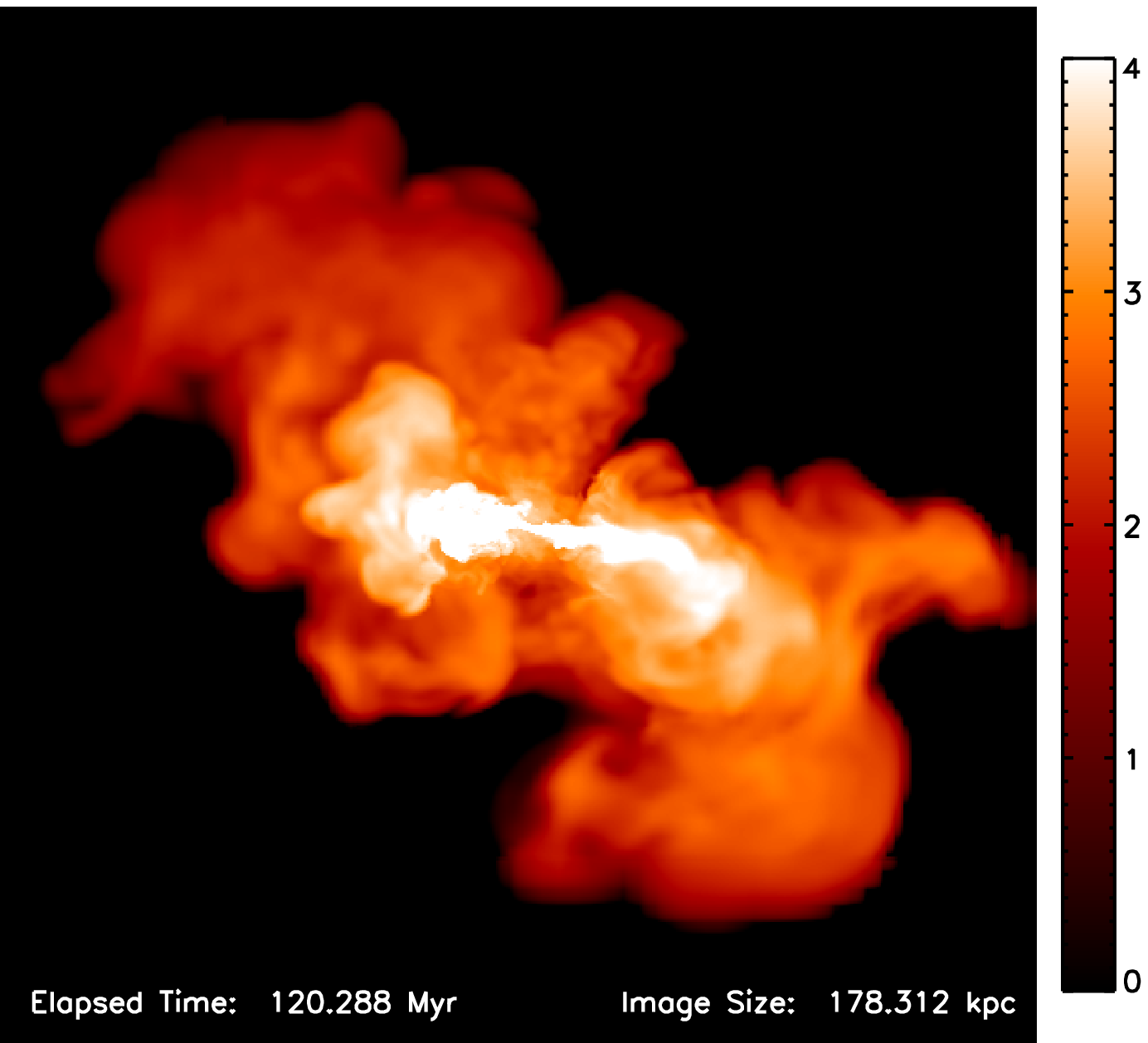}

\includegraphics[scale=.6]{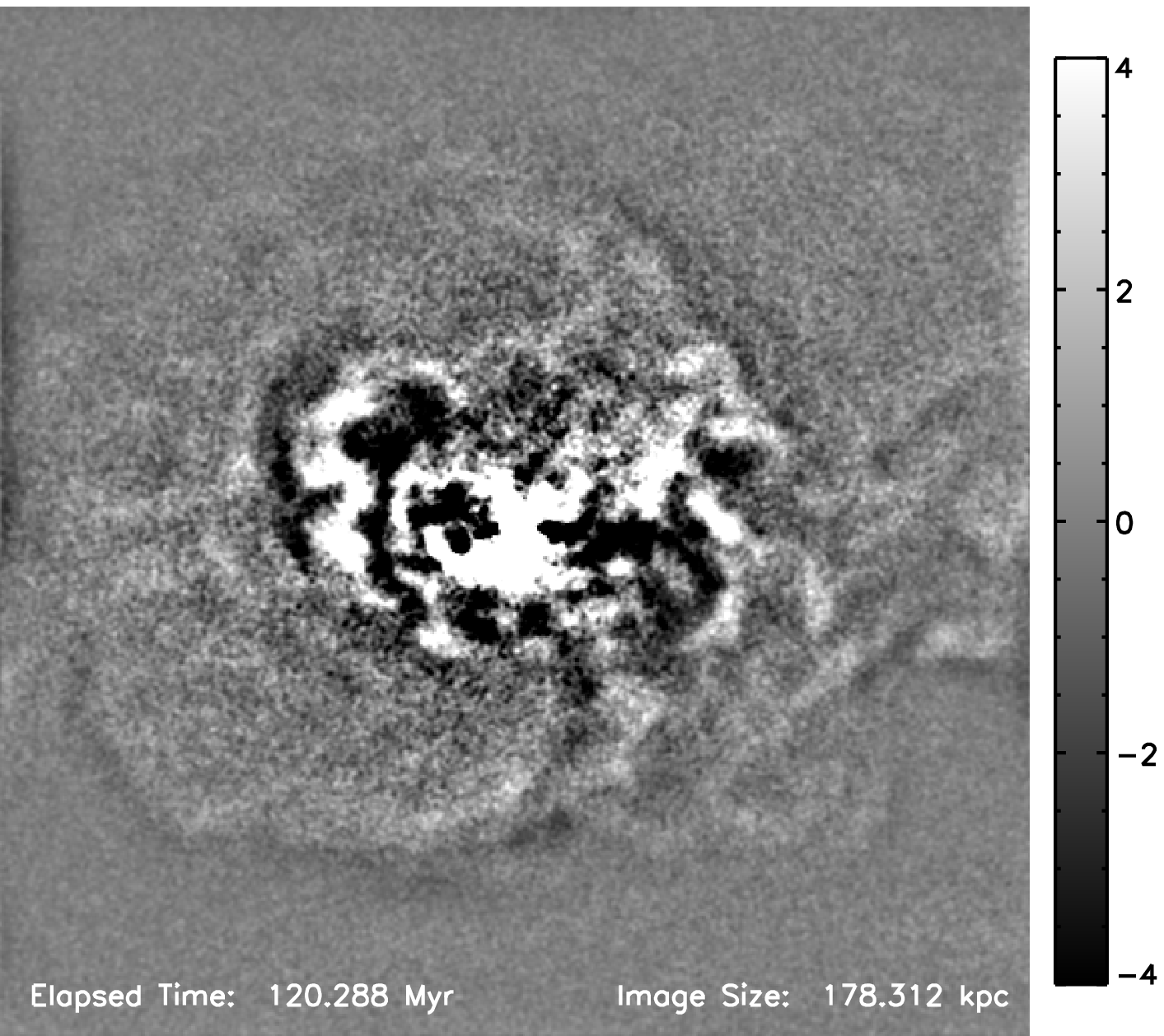}
\includegraphics[scale=.6]{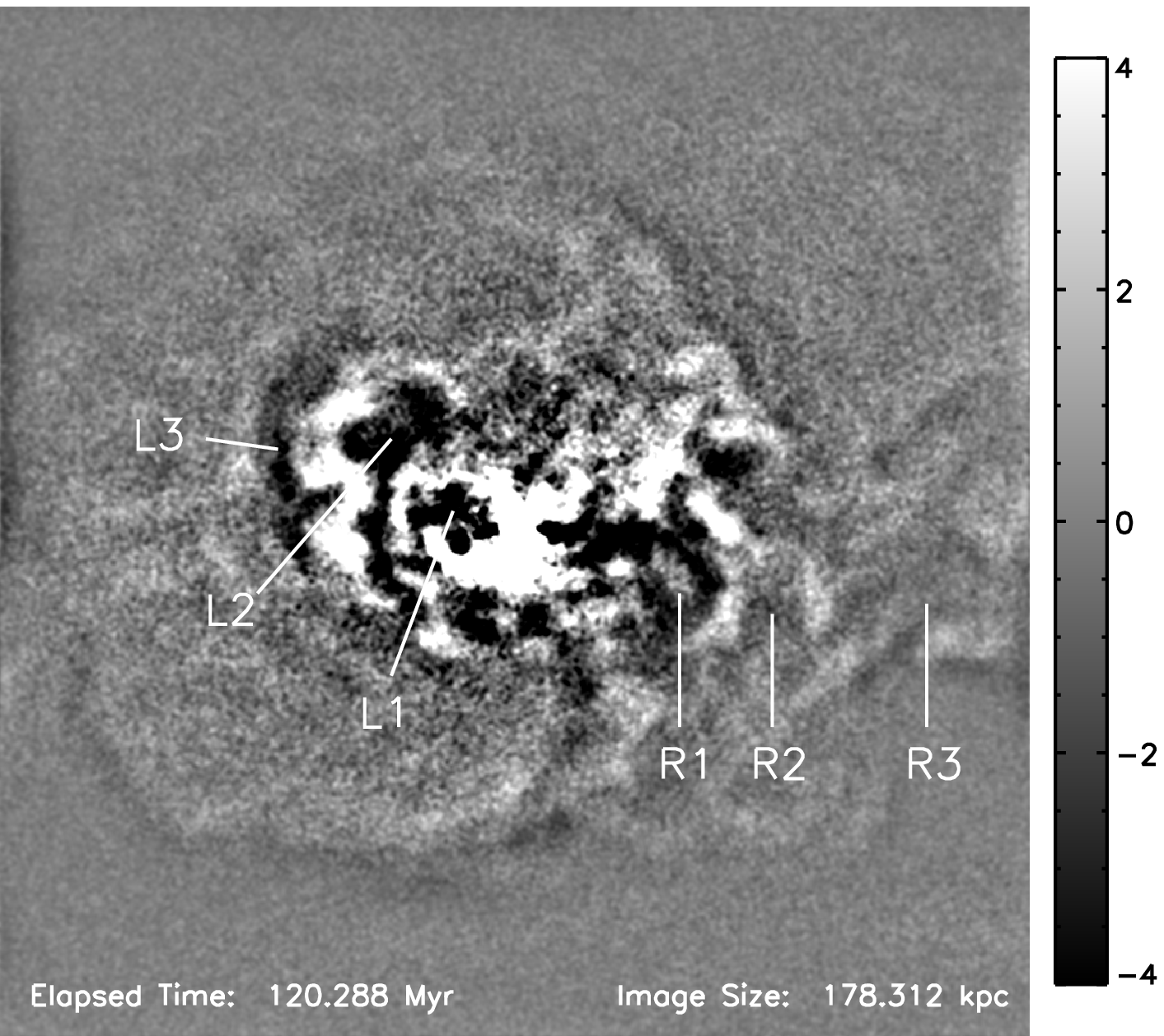}
\caption{
     Synthetic Chandra X-ray data (upper left, log scale, counts/pixel) and radio data (upper right, log scale, arbitrary units) for simulation with continuous AGN of $10^{45}$~erg/s (45C) after 120 Myr, at the distance of the Perseus cluster.  
     Lower left and lower right panels are an unsharp-masked image (with and without labels) of the X-ray data produced by the same procedure as in Fabian et al. 2003.  
     A series of bubbles detached from the AGN are visible to the upper left and lower right of the cluster centre, and are labelled L1 - L3 and R1 - R3 in the lower right image.
     Low level radio emission extends beyond the distinct bubbles visible in the X-ray images, although there are small ripples in the unsharp-masked image throughout the radio region.     
}
\label{fig_image_xray}
\end{figure*}

Our second group of simulations all have the same AGN luminosity of
$10^{45}$~erg/s, but the AGN is active for different amounts of time:
30~Myr for simulation 45, 90~Myr for 45L and
continuously until the end of the simulation at 200~Myr for 45C.  The
long-duration simulations begin identically to model 45 with the
inflation of a pair of bubbles that are deflected by the large-scale
circular motion of cluster gas.  Eventually, the bubbles are sheared
off and become disconnected from the jet.  At this point the bubble 
is no longer powered, but continues rise and be 
pushed by the cluster gas, which moves the bubble away from the jet axis.
The jet then inflates a second bubble behind the first which grows until it too
is broken off and advected away. 
The cycle of bubble inflation and break off leads to the existence of
multiple X-ray bubbles within the cluster.  In simulation 45C, there
are about 12 generations of bubble formation in 200 Myr.
Figure~\ref{fig_image_xray} shows a synthetic Chandra X-ray image and 
synthetic radio image for simulation 45C after 120 Myr, produced in 
the same way as the images in Fig.~\ref{fig_image_45}.
In the X-ray image (upper left panel) there is an inner bubble just to the
left of the cluster centre (extending to about 20 kpc) and a second,
outer bubble visible to the upper left of the cluster centre
(extending to about 40 kpc).  These two bubbles are also visible as 
distinct structures in the radio image (upper right panel).

Close examination of an unsharp-masked image (lower panels) shows these 
two bubbles (L1 and L2 in lower right panel) as well as a third (L3)
immediately to the left of the second.  All three of these bubble were 
formed by the AGN jet directed to the left.  On the right, there is a 
fairly large bubble reaching 40 kpc (R1), outside of which there is a 
second possible bubble reaching 65 kpc (R2), and then a third reaching 
about 85 kpc (R3).  Due to the low contrast between the interiors of 
bubbles R2 and R3 and their surroundings, they would likely be 
identified as waves rather than bubbles in an X-ray observation 
(see also En{\ss}lin \& Heinz (2002) for discussion of X-ray detectability 
of bubbles).

The radio image reveals an extended radio halo that goes beyond the 
bubbles identifiable in the X-ray images.  
However, there are several small ripples in the unsharp-masked images 
through the halo region.
Morphologically, the radio image appears similar to observations of 
the extended radio emission of M87 (Owen et al. 2000), 
with a large radio bubble containing smaller, highly-structured lobes, 
which are in turn powered by narrow jets from the AGN.

The unsharp-mask image in the right panel of
Fig.~\ref{fig_image_xray} and Fig.~3 from Fabian et al. 2003, produced
using the same technique, both show multiple bubbles and sound waves with a
similar size scale and morphology.  Clearly, in this case any
inference about the AGN duty cycle from the observation of sound waves
or multiple generations of bubbles would be erroneous.

\begin{figure*}
\includegraphics[scale=1.]{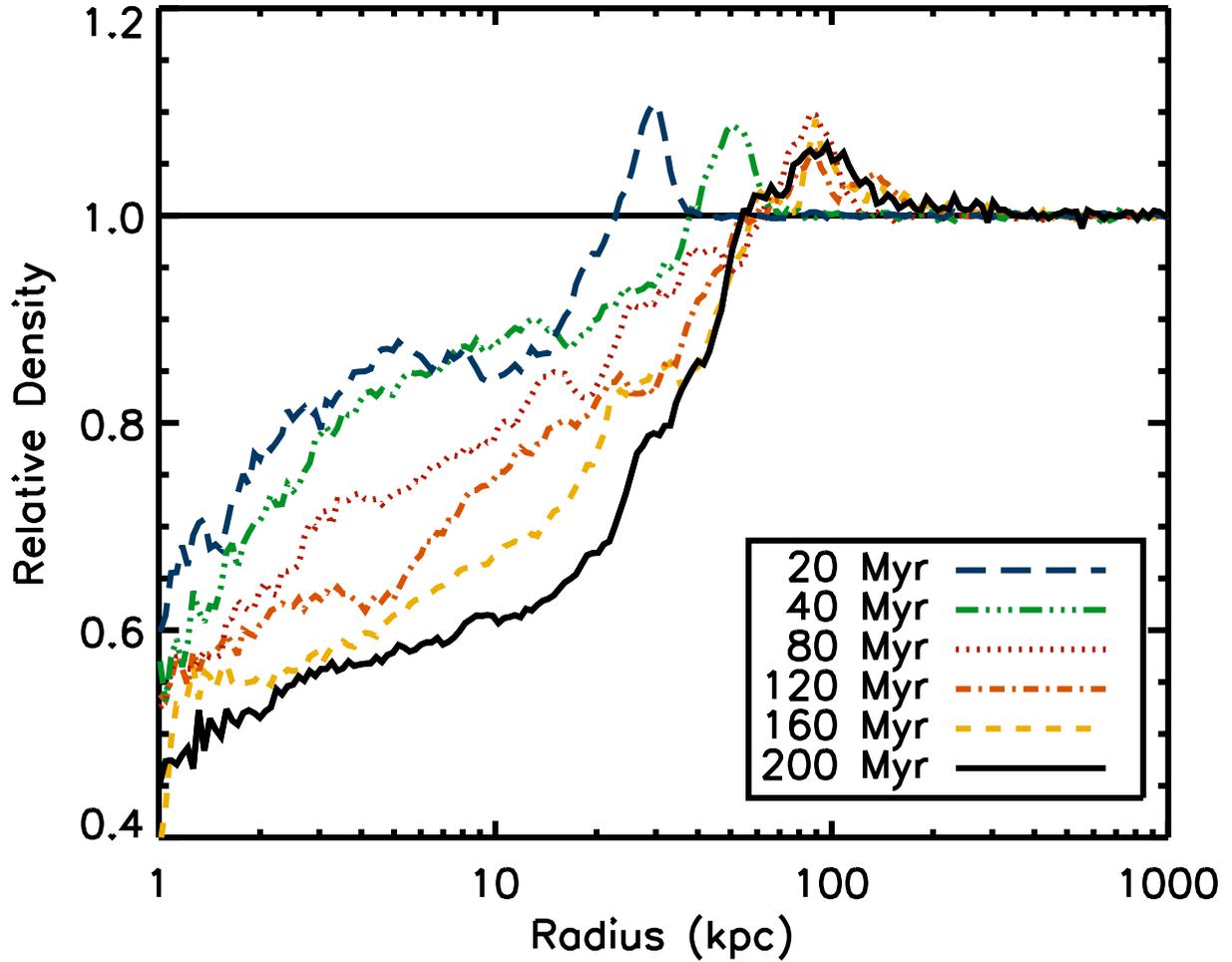}
\caption{
    Density relative to simulation with no AGN vs. radius for simulation 45C at from 20 to 200 Myr.
    After initially expanding, the radius of reduced density remains nearly constant at between 50 and 60 kpc after 80 Myr.
    However, the relative density continues to decrease, from about 75\% at 80 Myr to 55\% at 200 Myr.
}
\label{fig_dens_time_xlong}
\end{figure*}

The break off of bubbles also limits the radius affected by the AGN.
Figure~\ref{fig_dens_time_xlong} shows a time sequence of the average
density relative to a control simulation with no AGN vs.~radius for
simulation 45C.  The central density decrease expands to a radius of
about 60 kpc by 80 Myr, but then the radius remains fixed at between
50 and 60 kpc, even though the AGN continues to inject energy.  The
density in the centre of the cluster continues to drop with time as
more material is pushed out, but the material continues to pile up
between 50 kpc and 250 kpc without increasing the size of the region
being heated.

\begin{figure*}
\includegraphics[scale=1.]{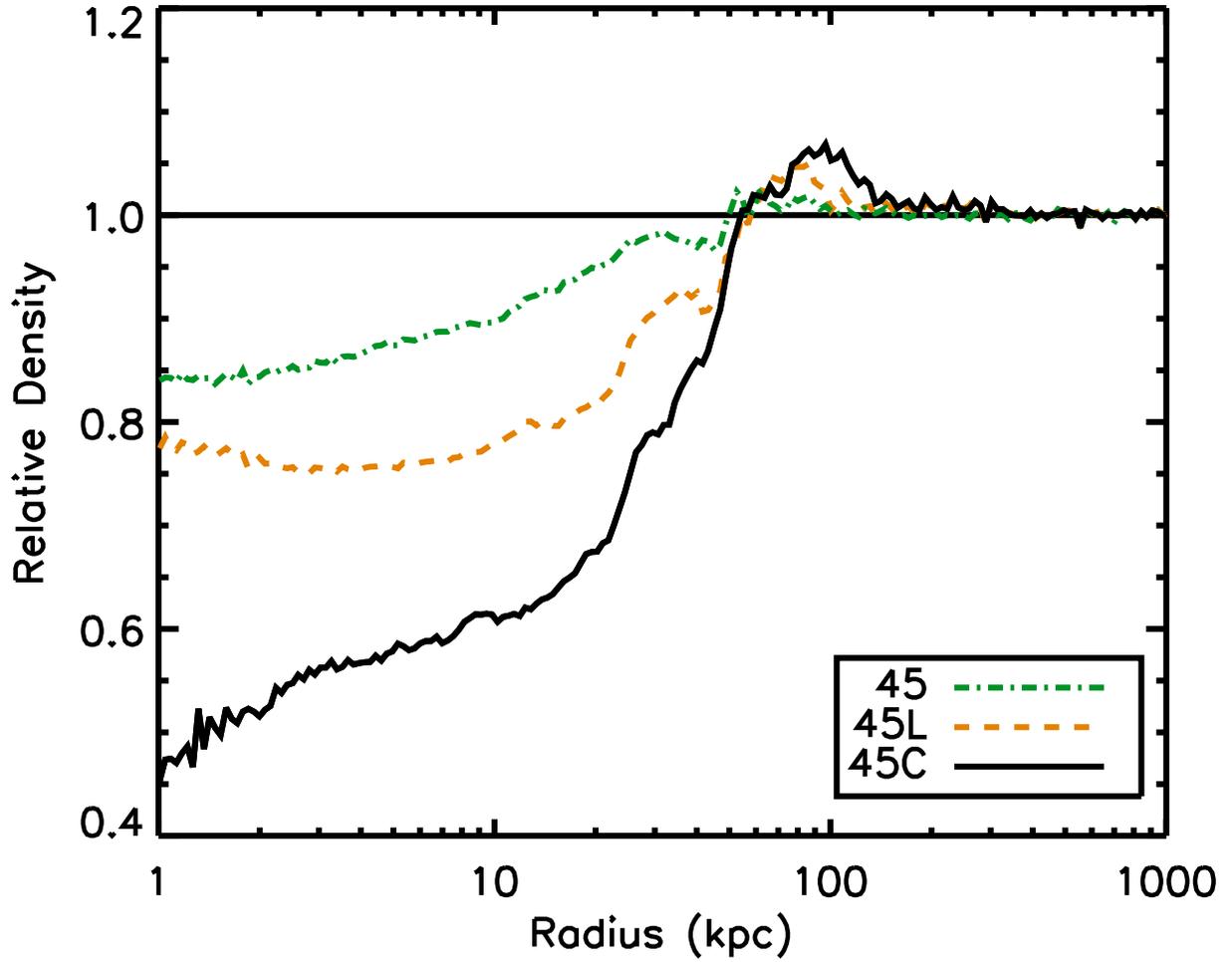}
\caption{
    Density relative to simulation with no AGN vs. radius after 200 Myr for simulations 45, 45L and 45C.
    For all three simulations the density has been decreased within a radius of about 50 kpc.
    The relative density is about 85\%, 75\% and 55\%, respectively, indicating that more material is removed from the centre of the cluster as the amount of energy injected increases.
}
\label{fig_dens_45}
\end{figure*}

Figure~\ref{fig_dens_45} compares the relative density of simulations
45, 45L and 45C after 200 Myr.  In all three cases the under-dense
region extends to 50 to 60 kpc with an over-dense region extending to
250 kpc.  The amount of material removed increases with AGN duration,
with the density decreasing in the inner few kpc by about 15\% for
simulation 45, 25\% for 45L and 40\% for 45C.  The density decrease
for simulation 45L is nearly the same as simulation 46
(Fig.~\ref{fig_dens_44_45_46}), although only 30\% as much energy has
been injected.  

This indicates that the impact made by a central AGN depends not only
on how much energy is injected but how it is injected.  The jet
luminosity determines the radius heated, while the amount of heating
is determined by a combination of luminosity and duration.  A
high-luminosity jet can reach a larger radius, but a low-power,
long-duration jet is more efficient at heating the centre of the
cluster.

\begin{figure*}
\includegraphics[scale=1.]{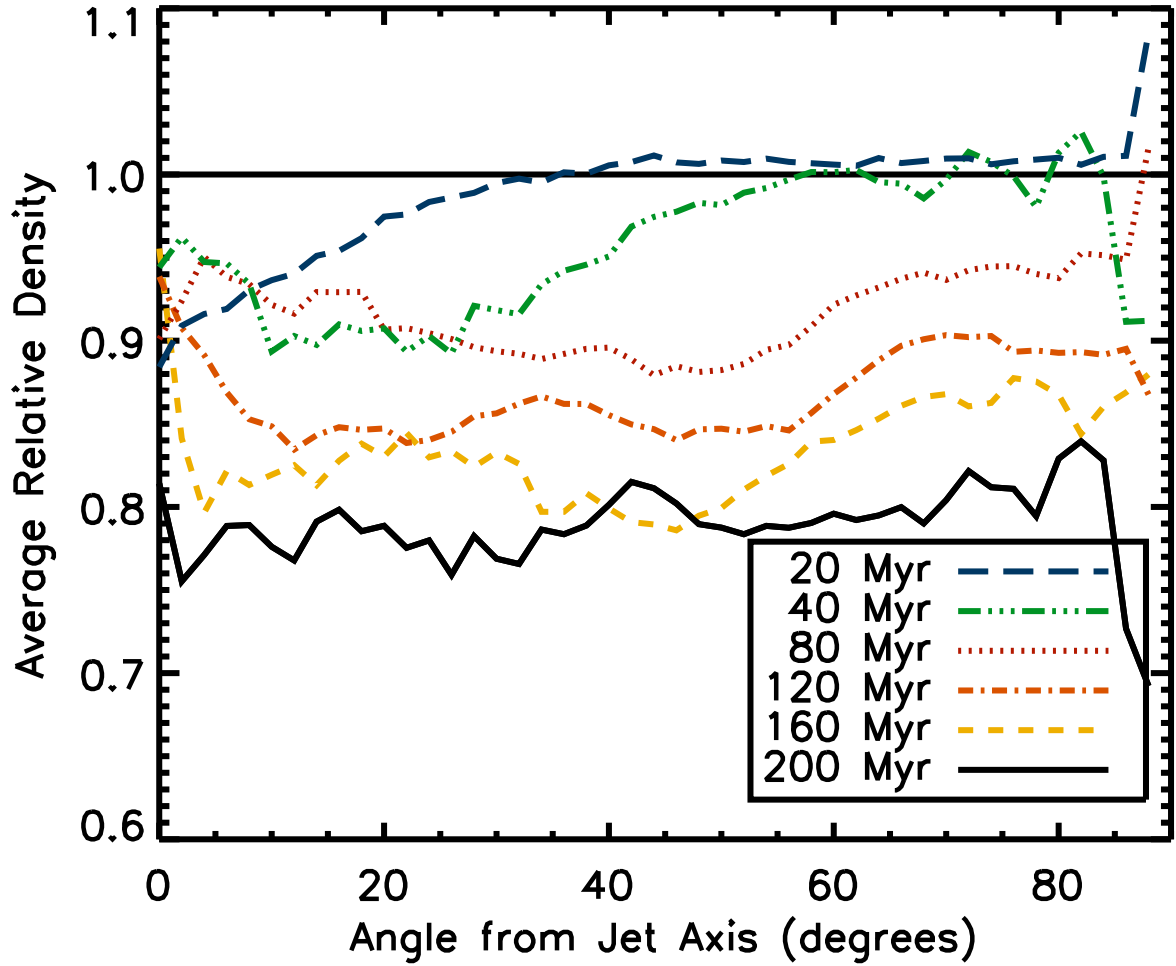}
\caption{
    Average density within 50 kpc relative to simulation with no AGN vs. angle for simulation 45C from 20 to 200 Myr.
    The effects of the AGN quickly spread over all angles and, by 80 Myr, the density distribution is uniform.
    Average density continues to decrease throughout the simulation, but the angular distribution is flat.
}
\label{fig_dens_time_angle_xlong}
\end{figure*}

The density decrease is still well distributed in angle in all three
simulations.  Figure~\ref{fig_dens_time_angle_xlong} plots the average
relative density within 50 kpc for model 45C from 20 to 200 Myr.  By
80 Myr, the density decrease is even distributed over all angles and
remains so even as the density continues to decrease.

\subsection{Comparison with Hydrostatic Simulation}

To verify that it is the motion of gas within the cluster that
determines the energy distribution, we have carried out one simulation
in a spherically symmetric hydrostatic cluster for comparison.  The
hydrostatic cluster has similar initial spherically averaged mass and
density profiles, but the radial pressure profile is adjusted to obey
hydrostatic balance. 
The result is a central temperature about $10\%$ higher than in the dynamic 
cluster to compensate for the lack of any rotational support.  
For simulation 45S, we injected jets in the cluster
centre with a luminosity of $10^{45}$~erg/s for the entire 200 Myr of
the simulation.  This is identical to the setup for simulation 45C,
aside from the different cluster environment.

\begin{figure*}
\includegraphics[scale=.6]{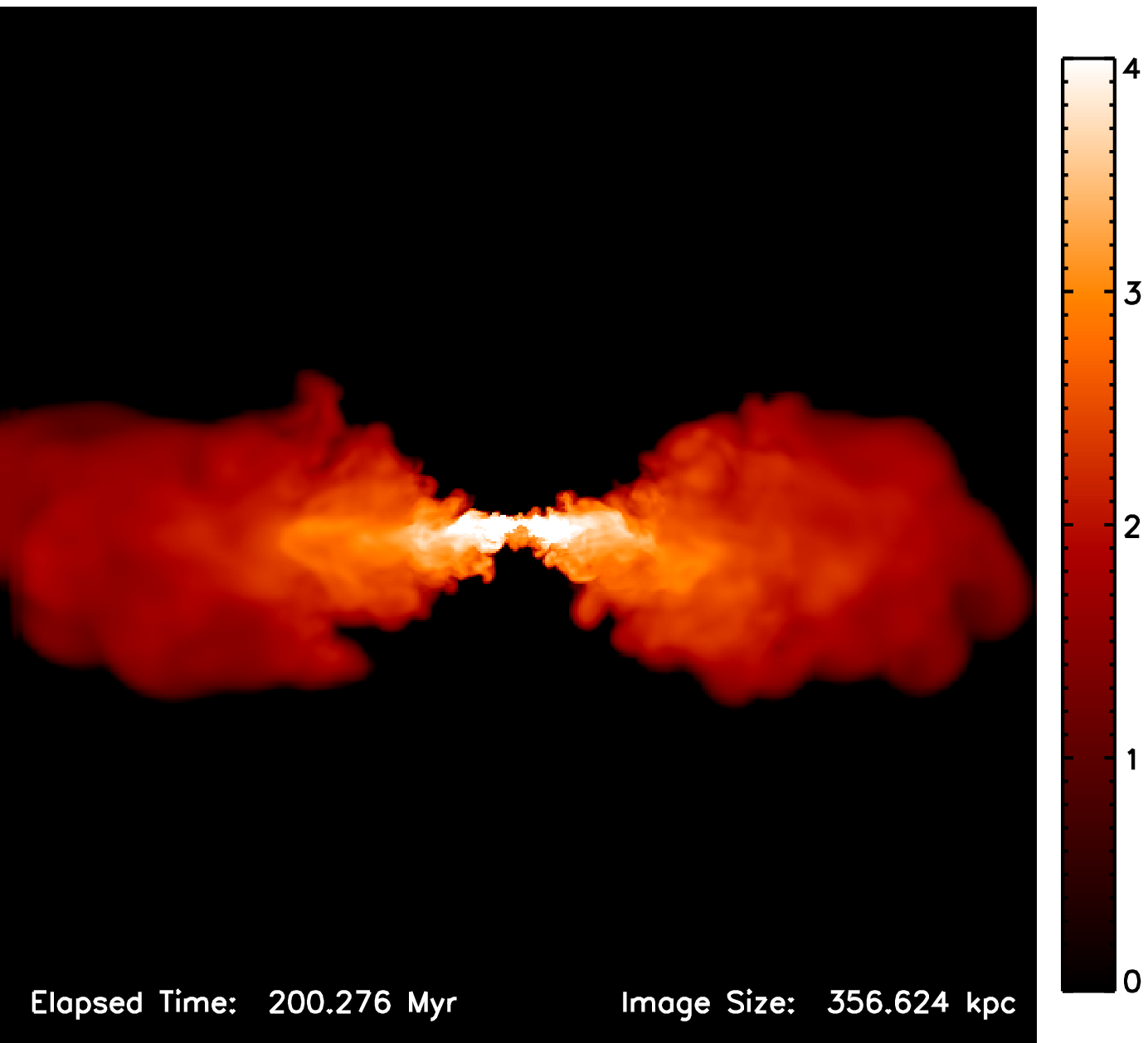}
\includegraphics[scale=.6]{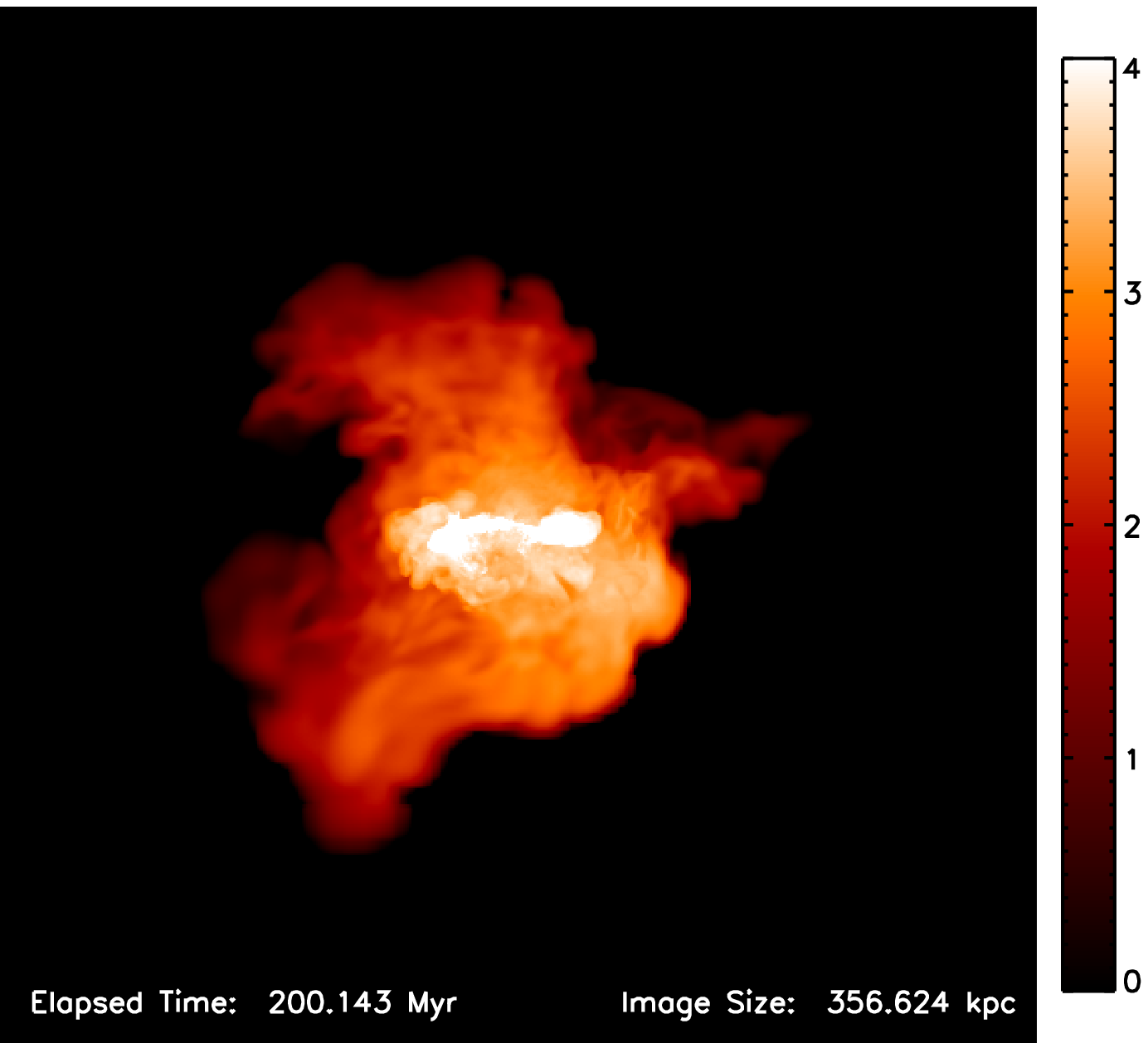}
\caption{
    Comparison of synthetic radio image (log scale) at 200 Myr for continuous AGN of luminosity $10^{45}$ erg/s for hydrostatic (left) and realistic (right) cluster (models 45S and 45C).  
    Without large scale flows, jet material in the hydrostatic cluster reaches a larger radius and remain close to the jet axis.
}
\label{fig_image_hydro}
\end{figure*}

A morphological comparison of synthetic radio emission from simulation
45S and 45C at 200 Myr (Fig.~\ref{fig_image_hydro}) shows clear
differences between the two cluster setups.  In the hydrostatic case
(left panel), the two large radio bubbles have been inflated to the
left and right of the cluster centre, aligned with the jet axis.  Most
jet material is within 20 degrees of the jet axis.  In the realistic
cluster, jet material is spread in all direction around the cluster
centre.  Jet material has also propagated farther from the cluster
centre, out to a maximum of about 200 kpc, rather than a maximum of
about 150 kpc in the realistic cluster.  This indicates that it is the
motion of gas in the cluster that spreads AGN material in angle and
limits it in radius.

\begin{figure*}
\includegraphics[scale=1.]{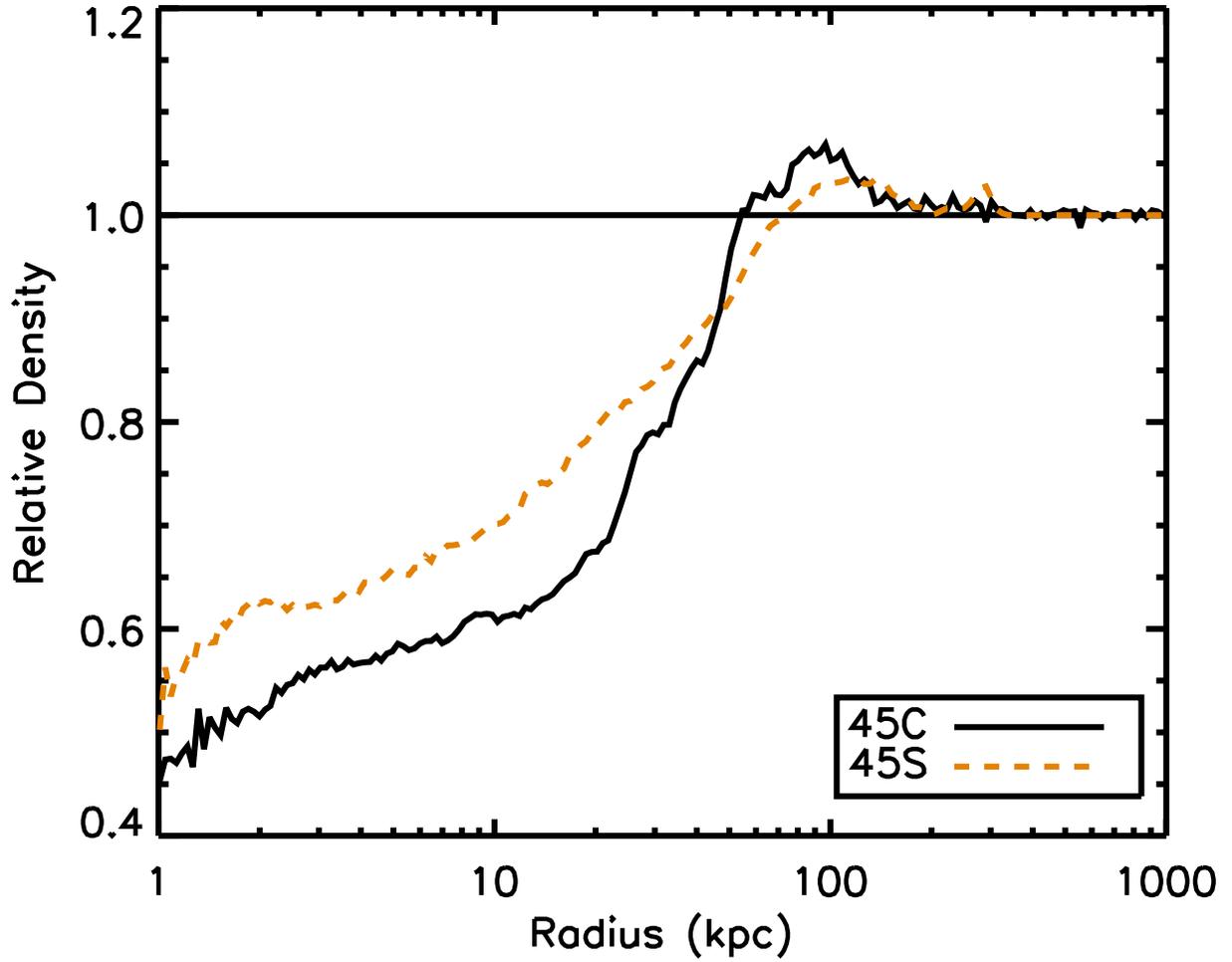}
\caption{
    Comparison of density relative to simulation with no AGN vs. radius after 200 Myr for continuous jets in hydrostatic and realistic clusters (models 45S and 45C).
    In the hydrostatic cluster density has been decreased out to 80 kpc rather than 50 kpc and there is some increase in density out to 300 kpc rather than 250 kpc.
}
\label{fig_dens_compare_static}
\end{figure*}

The differences between a static and a dynamic cluster can also be
seen in the density profiles of the two simulations.
Figure~\ref{fig_dens_compare_static} plots the relative density
vs. radius at 200 Myr for simulations 45C (realistic cluster) and 45S
(hydrostatic cluster).  The density decrease in the hydrostatic case
extends to about 80 kpc, rather than 50 kpc, and the over-dense region
reaches 300 kpc, rather than 250 kpc.

\begin{figure*}
\includegraphics[scale=1.]{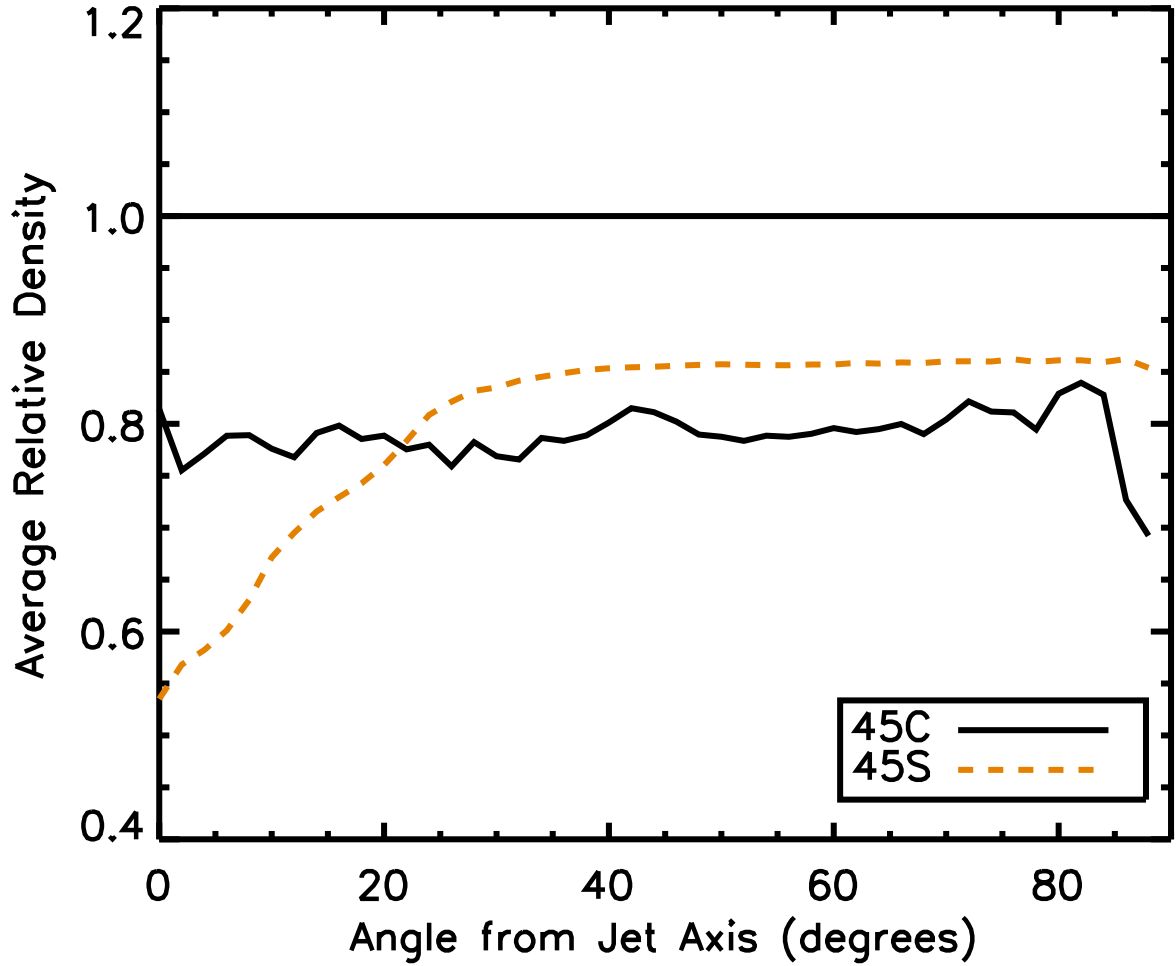}
\caption{
    Comparison of average density within 50 kpc relative to simulation with no AGN vs. angle after 200 Myr for hydrostatic and realistic clusters (models 45S and 45C).
    The density decrease is evenly distributed in a realistic cluster, but much more concentrated near the jet axis in a hydrostatic cluster.
}
\label{fig_dens_compare_angle_static}
\end{figure*}

Figure~\ref{fig_dens_compare_angle_static} plots the average relative
density within 50 kpc for the two simulations at 200 Myr.  While the
hydrostatic cluster simulation does have a density decrease at all
angles, the effect is about a factor of 2 stronger near the jet axis
compared to 30 degrees off axis, whereas the dynamic cluster
simulation has a uniform decrease at all angles.

\section{Analytic Model}

The scaling of radius with luminosity and the insensitivity of 
radius to jet duration can be understood with an analytic toy model 
that produces a characteristic timescale for bubble formation.
The jet initially inflates an expanding cocoon with the internal 
pressure balancing the ram pressure of the expanding shock, 
$p_c(t) = \rho(r_s) {\dot{r}_s}^2$, where $r_s$ is the radius of 
the external shock.

Following Kaiser \& Alexander (1997) and Heinz et al. (1998), for a 
density profile of the form 

\begin{eqnarray}
\rho = \rho_0 \left(\frac{r}{r_0}\right)^{-\alpha}
\end{eqnarray}

\noindent where $\rho_0$ is the density at radius $r_0$, there is a self-similar 
solution for the radius of the shock

\begin{eqnarray}
r_s = r_0 \left(\frac{t}{t_0}\right)^\frac{3}{(5-\alpha)}
\end{eqnarray}

\noindent where

\begin{eqnarray}
t_0 = C_1 \times \left(\frac{\rho_0 {r_0}^5}{L}\right)^{1/3}
\end{eqnarray}

\noindent and $t_0$ is the time at which $r_s = r_0$, $L$ is the jet luminosity 
and $C_1$ is a constant taken from Heinz et al. (1998).
The pressure in the cocoon then takes the form

\begin{eqnarray}
p_c(t) = \left(\frac{3}{(5-\alpha)}\right)^2 \rho_0 {r_0}^{2} {t_0}^{-2} \left(\frac{t}{t_0}\right)^\frac{-4-\alpha}{(5-\alpha)} \label{eqn_pc}
\end{eqnarray}

\noindent which decreases with time as $p_c \propto t^{\frac{-4-\alpha}{(5-\alpha)}} L^{\frac{2-\alpha}{(5-\alpha)}}$.
Because there is a circular motion in our cluster, there is 
an additional component of ram pressure on one side of each jet. 
In the inner 10 kpc of the cluster, density scales approximately as 
$\rho \propto r^{-1/2}$ and velocity scales as $v \propto r^{1/4}$, so 
the ram pressure due to circular motion is roughly constant, with a 
value of $p_{circ} = 3\times10^{-10}$~dyne/cm$^{2}$.
When the pressure in the cocoon drops below this value, the base of 
the cocoon will collapse and expanding bubble will be cut off from the jet.
Setting $p_c$ from Eqn.~\ref{eqn_pc} equal to $p_{circ}$, we can find the time 
for which the bubble will be cut off.
Outside the cluster centre, the density can be approximated by a power law 
with $\alpha = 1.5$ and $\rho_0 = 10^{-23}$~g/cm$^{3}$ for $r_0 = 1$~kpc
Using these values, we find a cutoff time, $t_{cut}$ of 

\begin{eqnarray}
t_{cut} \approx 33 \times L_{45}^{1/11}~Myr
\end{eqnarray}

\noindent where $L_{45} = L/(10^{45}$~erg/s$)$.  This time corresponds to a shock radius of

\begin{eqnarray}
r_{cut} \approx 22 \times L_{45}^{4/11}~kpc
\end{eqnarray}

In our simulations, the cutoff time appears to be about 20~Myr and the 
size of the cocoon at the cutoff time is about 40~kpc for the $10^{45}$~erg/s 
simulations, but for a toy model using a very simple density structure this is 
a reasonably close prediction.
Once the bubble is cut off, no more energy is added and it continues to expand 
and evolve independent of continued jet activity.
The cutoff time scales very weakly with luminosity ($L^{1/11}$), so the energy 
in the bubble when it is cut off will scale as $E_{bubb} \propto L^{12/11}$.
The radius at the cutoff scales as $L^{4/11}$, quite close to the $L^{\sim1/3}$ 
seen in our simulations.

After cutoff, the shock will continue to expand but without additional 
energy input.  
If the expansion remains supersonic, it will follow a Sedov-Taylor solution for an expanding shock, with 
$r_{s} \propto E^\frac{1}{(5-\alpha)} t^\frac{2}{(5-\alpha)}$ which is $\propto t^{4/7} L^{24/77}$ in this case (Chevalier 1976).
This is roughly consistent with the continued slow expansion seen in our 
simulations and is again close to $L^{1/3}$.

After the first bubble breaks off, the centre of the cluster is filled in with 
dense material and a new cocoon must form.
If the jet remains active, the new cocoon will go through the same process of 
growth and breakoff as the first.
Assuming the first bubble is advected away from the jet axis, the second bubble 
will be expanding into a density gradient similar to the first, so the time and 
size scale for the second breakoff will be similar as well.
The energy in the first bubble sets the maximum radius reached in the simulation, 
with subsequent bubbles following a similar expansion history.
Cutting off the bubble near the base of the jet provides an explanation for 
multiple generations of bubble formation with a timescale determined by the 
cluster parameters.

In our simulations, the radius of density decrease (and entropy increase) 
appears to approximately correspond to the radius of the shock at the time the 
first bubble breaks off.
Although disconnected bubbles rise beyond this radius, it may be that the 
strongest effects are limited to the region where the bubbles are still powered.

\section{Summary and Conclusions}

We have carried out a series of high-resolution hydrodynamic
simulations of AGN jets in the centres of galaxy clusters. 
Simulations are carried out in a realistic, dynamic cluster 
and cover a range of AGN luminosities and durations.

We find that the interaction of AGN jets with the motion of cluster
gas is critical for determining the evolution of both the cluster and
radio lobes.  In particular, we find that:

\begin{itemize}
\item Multiple bubbles can be formed from a single period of AGN
  activity with a constant luminosity.  As the AGN develops, bubbles
  can be broken off from the jet by the motion of gas within the
  cluster, leading to many generation of bubble formation.  These
  bubbles can be pushed away from the jet axis by cluster weather.
  Therefore, observations of multiple X-ray or radio bubbles in a
  cluster does not necessarily give you any information about the duty
  cycle of the AGN or about the past alignment of the AGN jet.

\item A toy model balancing pressure in an expanding cocoon against 
  ram pressure due to circular motion in the cluster core provides a 
  reasonable estimate for the timescale of bubble breakoff.

\item Energy from the AGN is distributed over all angles by
  large-scale motions in the cluster.  Jet material is distributed
  throughout the centre of the cluster and any information about the
  original orientation of the jet is lost on a time scale of about 100
  Myr. This is not the case in a hydrostatic cluster, where the
  absence of large-scale flows allows the jets to propagate without
  being deflected.  
  

\item A jet of a given luminosity will create a low-density, 
  high-entropy cavity that expands to a fixed radius and
  then stops, limited by the interaction with large-scale flows in the
  cluster.  The radius reached scales approximately as $R \propto L_{jet}^{1/3}$.  The
  exact radius reached is likely to depend on the detailed velocity
  and density structure within the cluster, but it does indicate that
  only high-luminosity AGN will be able to directly heat gas at 100
  kpc or more from the cluster centre.

\item How an AGN delivers its energy determines where that energy ends
  up in the cluster.  Lower luminosity AGN that are active for long
  periods are more efficient at heating the inner few 10's of kpc,
  while high luminosity AGN are necessary to deliver energy to large
  radii.  In our simulations, an AGN with a luminosity of
  $10^{45}$~erg/s that was active for 90 Myr was as
  effective at removing mass and increasing entropy in the inner 30
  kpc of the cluster as a $10^{46}$~erg/s AGN active for 30 Myr,
  despite emitting only 30\% as much total power.  However, the
  effects of the $10^{45}$~erg/s AGN were limited to half the radius of
  the $10^{46}$~erg/s AGN.

\item In a hydrostatic cluster, AGN evolution is quite different.  A
  long-duration AGN inflates two large bubbles rather than many
  smaller ones.  Jet material remains concentrated near the jet axis
  and the radius reached by the jet material continues to increase
  with time beyond the value we find in the dynamic case.

\end{itemize}

The relationship between jet power and the radius of the jets ``sphere
of influence'' has consequences for the impact jets have on clusters.
The strong effect that the motion of cluster gas has on the AGN
development means that exactly how and where an AGN deposits energy
will be strongly affected by the inflation history of the jet and the
dynamical state of the cluster.  


Low-power AGN are more efficient at heating the central cluster, but 
the heat is confined, implying that an additional heat source may be 
needed farther from the centre of the cluster.
One possible heat source is conduction of heat from warm gas in the 
outer cluster to cooling gas toward the centre.
Recent work by Parrish et al. (2010) and Ruszkowski \& Oh (2010) has shown that turbulence can 
suppress the heat-flux-driven buoyancy instability and allow efficient 
thermal conduction to occur.
AGN activity can act as a source of turbulence, potentially providing a 
switch that allows conduction to occur when an AGN is active.
Heinz et al. (2010), using some of the same simulations presented here, 
found an increase in the turbulent velocity dispersion of cluster gas 
due to AGN activity of up to several $100$~km~s$^{-1}$.


Central heating in cooling flows (where the entropy has to be injected
in a relatively small volume) is more likely to result from
continuously operating lower power jet than episodic powerful
outbursts.  This is, in a sense, numerical confirmation of the
effervescent model, where continuous low-level activity generated
multiple generations of bubbles (Ruszkowski \& Begelman 2002).  
However, in our case, the effect of
cluster weather is primarily responsible for breaking off bubbles, not
necessarily the bubble's buoyant escape.

The initial structure of a cluster will play a large role in determining 
the specific morphology of an AGN outflow.
Although we only use one realistic cluster setup for our simulations, it 
is clear that the assumption of a simple one-to-one correspondence between 
observations of X-ray cavities or waves and periods of activity from an 
intermittent AGN is not valid.
Multiple bubbles could be formed by intermittent AGN activity, but 
it is possible to produce multiple generations of X-ray bubbles from a 
single, continuously active AGN.

The simulations presented here do not include magnetic fields.  
The presence of magnetic fields could stabilize the bubbles created by 
the AGN, allowing them to rise farther away from the cluster centre 
before becoming disrupted (Ruszkowski et al. 2007).  
Observations of an increase in X-ray cavity size with distance from 
the cluster centre have been interpreted as favoring a current-dominated 
MHD jet model (Diehl et al. 2008).
However, Br\"uggen et al. (2009) argued that pure hydro simulations could 
produce similar observations.
Even if the bubble evolution is not dominated by magnetic 
fields, the presence of weak magnetic fields will affect the mixing of jet 
and cluster material and can change where energy is deposited.
Future simulations including realistic magnetic fields will assess 
the effect of magnetic fields on cluster evolution.

\section*{Acknowledgments}

BM and SH acknowledge support from NSF grant 0707682.
MB acknowledges support from the DFG under grant BR 2026/3 within 
the Priority Programme ``Witnesses of Cosmic History.''
MR acknowledges support from NASA Chandra theory grant TM8-9011X.
The software used in this work was in part developed by the
DOE-supported ASC/Alliance Center for Astrophysical Thermonuclear
Flashes at the University of Chicago.


%
%
%

%
%
%
%


\label{lastpage}

\end{document}